%% file: CoMelSinger.tex
\documentclass[lettersize,journal]{IEEEtran}
\usepackage{amsmath,amsfonts}
\usepackage{algorithmic}
\usepackage{algorithm}
\usepackage{array}
\usepackage[caption=false,font=normalsize,labelfont=sf,textfont=sf]{subfig}
\usepackage{textcomp}
\usepackage{stfloats}
\usepackage{url}
\usepackage{verbatim}
\usepackage{graphicx}
\usepackage{booktabs}
\usepackage{amssymb}
\usepackage{dsfont}
\usepackage{xcolor}
\usepackage{multirow}
\usepackage{cite}
\usepackage[table]{xcolor}
\usepackage{subcaption}
\usepackage[colorlinks=true, linkcolor=blue, citecolor=blue, urlcolor=blue]{hyperref}
\usepackage{orcidlink}
\begin{document}
\bstctlcite{IEEEexample:BSTcontrol}
\title{CoMelSinger: Discrete Token-Based Zero-Shot Singing Synthesis With Structured Melody Control and Guidance}

\author{Junchuan Zhao\orcidlink{0009-0008-2616-6590}, Wei Zeng\orcidlink{0000-0002-0953-5314}, Tianle Lyu\orcidlink{0009-0004-4362-681X}, Ye Wang\orcidlink{0000-0002-0123-1260},~\IEEEmembership{Member,~IEEE}
\thanks{Junchuan Zhao, Wei Zeng, Tianle Lyu, Ye Wang, are affiliated with the School of Computing, National University of Singapore, Singapore. Ye Wang is the correspondence author of this paper. Contact junchuan@comp.nus.edu.sg for further questions about this work.}        % <-this % stops a space
}


\maketitle

\begin{abstract}
\input{sections/abstract}
\end{abstract}


\section{Introduction}
\input{sections/intro}\label{sec:intro}

\section{Related Works}
\input{sections/relat}

\section{Method}
\input{sections/method}

\section{Experimental Setups}
\input{sections/exp}

\section{Experimental Results}
\input{sections/result}

\section{Conclusion}
\input{sections/concl}

\bibliographystyle{IEEEtran}
\bibliography{Bibliography}

\end{document}

%% file: sections/abstract.tex
Singing Voice Synthesis (SVS) aims to generate expressive vocal performances from structured musical inputs such as lyrics and pitch sequences. While recent progress in discrete codec-based speech synthesis has enabled zero-shot generation via in-context learning, directly extending these techniques to SVS remains non-trivial due to the requirement for precise melody control. In particular, prompt-based generation often introduces prosody leakage, where pitch information is inadvertently entangled within the timbre prompt, compromising controllability. We present CoMelSinger, a zero-shot SVS framework that enables structured and disentangled melody control within a discrete codec modeling paradigm. Built on the non-autoregressive MaskGCT architecture, CoMelSinger replaces conventional text inputs with lyric and pitch tokens, preserving in-context generalization while enhancing melody conditioning. To suppress prosody leakage, we propose a coarse-to-fine contrastive learning strategy that explicitly regularizes pitch redundancy between the acoustic prompt and melody input. Furthermore, we incorporate a lightweight encoder-only Singing Voice Transcription (SVT) module to align acoustic tokens with pitch and duration, offering fine-grained frame-level supervision. Experimental results demonstrate that CoMelSinger achieves notable improvements in pitch accuracy, timbre consistency, and zero-shot transferability over competitive baselines. Audio
samples are available at \url{https://danny-nus.github.io/CoMelSinger/}.

%% file: sections/intro.tex
\IEEEPARstart{S}{inging} voice synthesis (SVS) aims to transform structured musical inputs—most often lyrics and pitch sequences—into expressive, high-quality vocal performances. Over the past decade, it has moved from a niche research topic to an essential tool in creative audio technologies, propelled by the rise of AI-driven music generation, virtual performers, and personalized media experiences. Its applications now extend well beyond traditional karaoke systems, finding a place in virtual idol production, game soundtracks, and content creation for social platforms. Parallel to these expanding use cases, advances in deep generative models have brought marked gains in timbre fidelity, pitch accuracy, and the overall naturalness of synthesized voices \cite{guo2025techsinger}, \cite{hono2021sinsy}, \cite{hwang2025hiddensinger}, \cite{liu2022diffsinger}, \cite{ye2023comospeech}, \cite{zhang2022visinger}, \cite{zhang2024stylesinger}, \cite{zhao2024sintechsvs}.

Recent SVS frameworks, including end-to-end~\cite{zhang2022visinger,zhang2022visinger2} and diffusion-based architectures~\cite{liu2022diffsinger,zhao2024sintechsvs,byun2025hierarchical,byun2024midi,hwang2025hiddensinger}, have demonstrated strong performance in supervised scenarios with seen singers. Nevertheless, zero-shot SVS \cite{huang2023make} \cite{zhao2025spsinger}, \cite{byun2025hierarchical}, \cite{dai2025everyone}—synthesizing singing voices for unseen speakers without additional fine-tuning—still has substantial room for improvement. 

\begin{figure}[t]
\centering
\includegraphics[width=1.0\columnwidth]{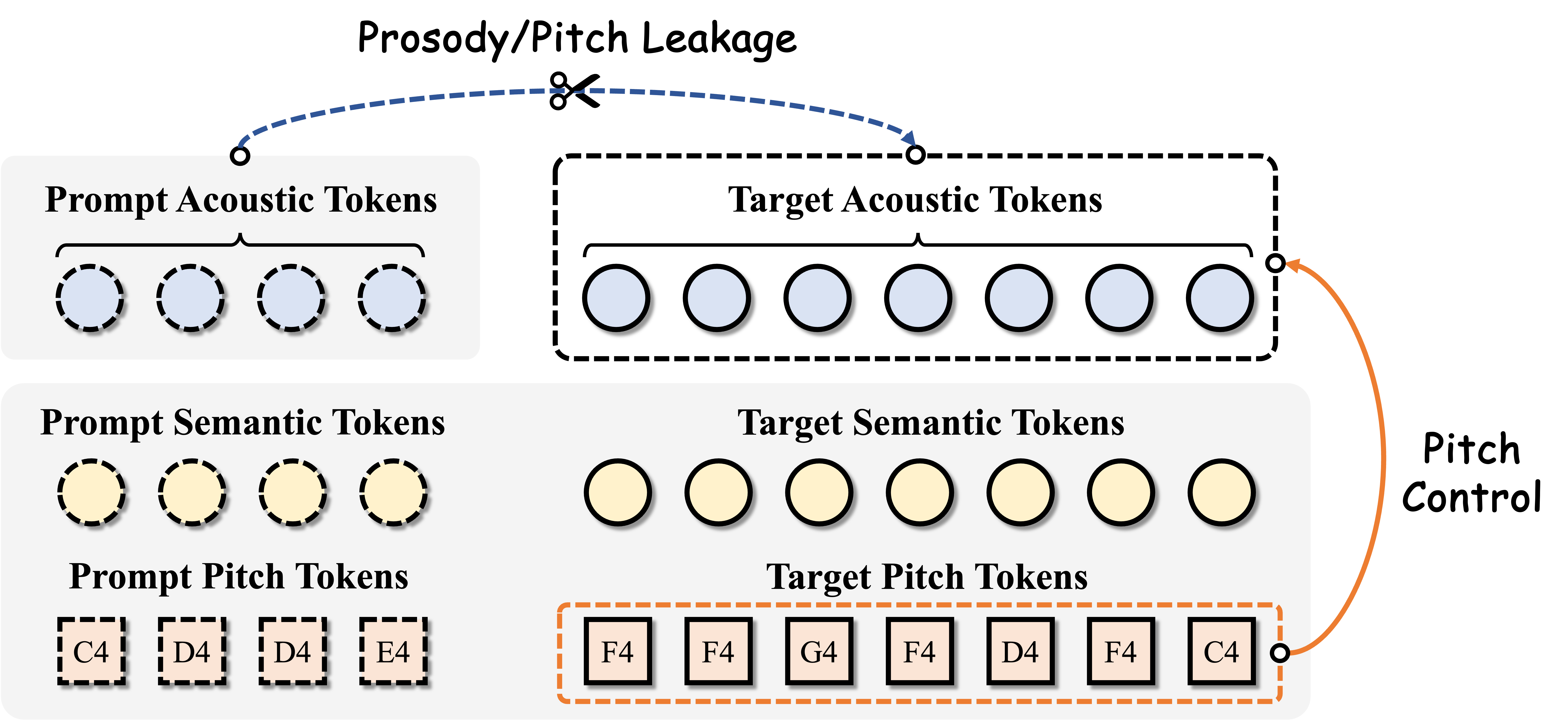} 
% Reduce the figure size so that it is slightly narrower than the column. Don't use precise values for figure width.This setup will avoid overfull boxes. 
\caption{Illustration of pitch leakage in prompt-based SVS. Despite conditioning on lyrics (semantic tokens) and pitch tokens, the model may still infer prosodic cues from prompt acoustic tokens, leading to pitch or prosody leakage.}
\label{fig:prosody_leakage}
\end{figure}

\IEEEpubidadjcol{Discrete} token-based architectures show great in-context learning capabilities and provide a promising pathway for such zero-shot generation. In a related task, text-to-speech (TTS) has witnessed rapid progress with discrete acoustic tokens derived from vector quantization and neural audio codecs~\cite{wang2023neural}, \cite{du2024cosyvoice}, \cite{du2024cosyvoice2}, \cite{wang2024maskgct}, \cite{ju2024naturalspeech}. By mapping complex waveforms into a quantized latent space, these tokens capture timbre, prosody, and phonetic content, thereby reformulating speech synthesis as symbolic sequence modeling akin to language modeling. The success of token-based TTS systems is largely enabled by large-scale multi-speaker corpora, which provide sufficient diversity to learn robust and generalizable representations. In contrast, the scarcity and limited diversity of singing data make token-based modeling for SVS significantly more challenging.

Built upon this data-rich foundation, recent TTS systems \cite{wang2023neural}, \cite{wang2024maskgct}, \cite{guo2024socodec}, \cite{borsos2023soundstorm}, \cite{du2024cosyvoice}, \cite{du2024cosyvoice2}, have adopted large language model (LLM)-style architectures to model the conditional distribution of acoustic tokens given phoneme sequences and optional prompts. Within this framework, in-context learning (ICL) becomes feasible: a short segment of reference speech, represented as discrete tokens, serves as an acoustic prompt to guide synthesis in terms of speaker identity and style. This approach, exemplified by models such as VALL-E \cite{wang2023neural}, enables zero-shot speech synthesis by treating speech generation as a form of conditional codec language modeling, without requiring speaker labels or model adaptation. Inspired by these advances, researchers have begun exploring discrete token-based methods for SVS.

However, directly extending TTS by replacing textual input with structured musical inputs—such as lyrics and pitch tokens—while reusing the same modeling pipeline reveals a unique challenge in the singing domain: prosody leakage from the acoustic prompt, as illustrated in Figure~\ref{fig:prosody_leakage}. In prompt-based synthesis, the acoustic prompt is intended to provide timbral cues, yet pitch-related attributes—including contour and timing—are often inadvertently encoded into its latent representation. This unintended encoding leads to timbre–melody entanglement, where the prompt simultaneously influences vocal timbre and melodic realization. As a result, the system’s control over the explicitly specified pitch sequence is weakened, undermining the precise separation between timbre conditioning and melody generation that SVS requires.

The disparity in dataset size between singing \cite{wang2022opencpop}, \cite{zhang2022m4singer}, \cite{dai2023singstyle111}, and speech \cite{zen2019libritts}, \cite{he2024emilia}, \cite{kahn2020libri}, further exacerbates the difficulty of addressing prosody leakage. Importantly, the effectiveness of in-context learning in TTS relies on large-scale, diverse speech corpora, which enable robust learning of disentangled token representations. In contrast, singing datasets are generally smaller and less varied, making it harder to avoid prosodic interference and to generalize prompt-based conditioning. Consequently, directly transferring token-based prompting strategies from TTS to SVS often leads to weaker melody control in zero-shot scenarios. In this case, achieving reliable melody control is particularly crucial for maintaining alignment with the musical score.

Moreover, SVS requires much finer control over pitch and melody than TTS or voice conversion, as the generated singing must accurately follow the musical score while preserving timbre. Although attribute control has been explored through adversarial training~\cite{ju2024naturalspeech}, contrastive learning~\cite{wu2024dctts}\cite{tang2022avqvc}, and information-bottleneck methods~\cite{qian2020unsupervised}\cite{zhao2022disentangling}\cite{lian2022robust}\cite{zhao2025prosody}, these approaches primarily focus on coarse prosodic patterns or emotional cues and provide limited support for fine-grained melody control. Make-A-Voice~\cite{huang2023make}, as a representative discrete-token SVS system, adopts prompt-guided conditioning but lacks explicit mechanisms to prevent the acoustic prompt from influencing melody realization. 
Recent unified speech-and-singing models such as Vevo 1.5\footnote{\url{https://github.com/open-mmlab/Amphion/tree/main/models/svc/vevosing}}/2.0 \cite{zhang2025vevo2} take a representation-level perspective by introducing melody- or prosody-aware tokenizers that explicitly model musical F0 structures, aiming to reduce melody leakage through improved disentanglement. However, these approaches typically rely on external melody audio (e.g., piano or humming recordings) as conditioning signals, which can introduce alignment ambiguities between melody and lyrics and limit the precision of fine-grained melody control.

To address the challenges of prosody leakage and limited melody controllability in zero-shot SVS, we propose CoMelSinger, a discrete codec-based framework with structured melody control.
CoMelSinger builds on the non-autoregressive MaskGCT architecture~\cite{wang2024maskgct}, adapting it to accept musical inputs consisting of lyrics and pitch tokens.
To achieve better melody–timbre control, we introduce a coarse-to-fine contrastive learning strategy that limits excessive pitch-related information in the acoustic prompt, allowing the explicit pitch condition to guide melodic realization more effectively.
We further incorporate a lightweight, encoder-only Singing Voice Transcription (SVT) module to provide fine-grained, frame-level supervision by aligning acoustic tokens with pitch and duration sequences.
Together, these designs enable accurate melody modeling, maintain timbre consistency, and preserve the in-context learning capability of discrete token-based systems.
Extensive experiments on both seen and unseen singers demonstrate that CoMelSinger delivers substantial improvements in pitch accuracy, timbre consistency, and overall synthesis quality compared with state-of-the-art SVS baselines in zero-shot scenarios. The main contributions of this work are:
\begin{itemize}
\item We propose CoMelSinger, a discrete token-based SVS framework for zero-shot synthesis with structured melody control.
\item We introduce a coarse-to-fine contrastive learning mechanism to improve melody–timbre control by limiting excessive pitch-related information in the acoustic prompt.
\item We develop a lightweight SVT module that aligns acoustic tokens with pitch and duration, providing frame-level supervision to improve melody fidelity.
\item Comprehensive experiments on public SVS datasets demonstrate that CoMelSinger achieves superior pitch accuracy, timbre consistency, and generalization to unseen singers.
\end{itemize}

%% file: sections/relat.tex
\subsection{Singing Voice Synthesis}\label{subsec:SVS}

\begin{figure}[t]
\centering
\includegraphics[width=\columnwidth]{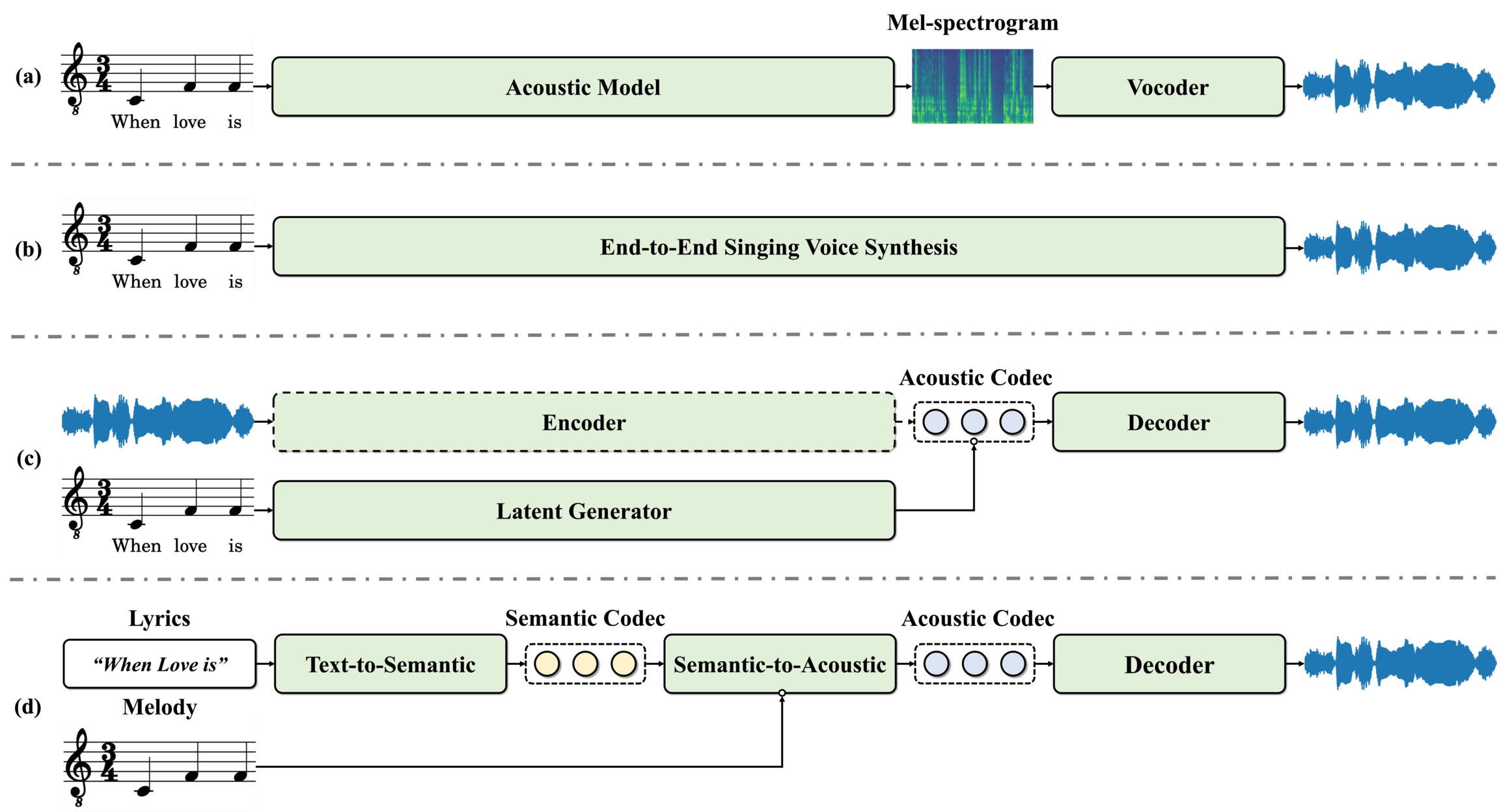} 
% Reduce the figure size so that it is slightly narrower than the column. Don't use precise values for figure width.This setup will avoid overfull boxes. 
\caption{Comparison of SVS system architectures. (a) Two-stage pipeline using a pre-defined continuous intermediate representation. (b) End-to-end system mapping directly from musical input to audio. (c) HiddenSinger-style \cite{hwang2025hiddensinger} system employing a pretrained audio codec, with frozen components during codec prediction and audio synthesis (dashed outlines). (d) Three-stage SVS systems such as Make-a-Voice \cite{huang2023make} and our proposed CoMelSinger, utilizing discrete intermediate representations.}
\label{fig:SVS_compare}
\end{figure}

Singing voice synthesis (SVS) aims to produce expressive vocal performances from structured musical inputs such as note pitch, duration, and lyrics.
Compared with text-to-speech (TTS), SVS presents unique challenges, including a wider pitch range and sustained phonation, which demand fine-grained melody modeling.
Figure~\ref{fig:SVS_compare} illustrates the evolution of representative SVS architectures, from continuous-feature pipelines to end-to-end models and, more recently, discrete codec-based frameworks.
Early SVS systems relied on unit-selection synthesis (e.g., VOCALOID~\cite{kenmochi2007vocaloid, bonada2007synthesis}) or statistical approaches based on hidden Markov models (HMMs)\cite{saino06_interspeech}.
With the advent of deep learning, SVS models increasingly adopted a two-stage architecture—comprising an acoustic model followed by a vocoder—including XiaoiceSing\cite{lu2020xiaoicesing}, DeepSinger~\cite{ren2020deepsinger}, and Sinsy~\cite{hono2021sinsy}.
Subsequent advances incorporated generative adversarial networks (GANs)\cite{chandna2019wgansing, huang2022singgan} to enhance timbre realism, while more recent work using denoising diffusion probabilistic models (DDPMs)\cite{liu2022diffsinger, zhao2024sintechsvs} has achieved further gains in fidelity and temporal coherence.
In parallel, end-to-end systems such as VISinger 1/2~\cite{zhang2022visinger, zhang2022visinger2} have been developed to generate waveforms directly from musical scores without relying on explicit intermediate features.

Inspired by the success of discrete token modeling in TTS~\cite{wang2023neural, wang2024maskgct, du2024cosyvoice}, recent SVS studies have explored token-based representations to improve generalization.
TokSing~\cite{wu2024toksing} employs a non-autoregressive Transformer conditioned on lyrics and pitch embeddings to predict discrete acoustic tokens.
HiddenSinger~\cite{hwang2025hiddensinger} integrates a diffusion-based decoder guided by discrete pitch and semantic tokens, achieving high-quality synthesis.
Make-A-Voice~\cite{huang2023make} unifies speech and singing synthesis through a shared discrete representation, but uses a relatively small proportion of singing data and does not incorporate prompt-based in-context learning.
Consequently, it lacks explicit melody conditioning and provides limited flexibility in zero-shot singing scenarios. 
Recently, models such as Vevo 1.5/2.0 \cite{zhang2025vevo2} have proposed addressing melody leakage through melody- or prosody-aware tokenizers that explicitly model musical F0 structures. While these methods mitigate melody leakage by redesigning the tokenizer, our work focuses on improving melody control within a prompt-based SVS framework by introducing explicit melody inputs and regulated prompt-based conditioning, enabling more accurate and controllable zero-shot singing voice synthesis.

\subsection{Discrete Token Based Speech Synthesis}\label{subsec:D-TTS} 
Discrete speech modeling has gained momentum following advances in self-supervised learning (SSL) for speech representation.
Models such as HuBERT~\cite{hsu2021hubert} and Wav2Vec 2.0~\cite{baevski2020wav2vec} learn meaningful latent representations from raw audio, which can be quantized into discrete units for downstream tasks.
These discrete units provide compact and controllable representations, enabling applications such as low-bitrate speech coding and voice conversion.

Inspired by the success of large language models (LLMs), recent approaches formulate speech synthesis as autoregressive generation over discrete codec tokens.
VALL-E~\cite{wang2023neural} pioneered this direction by conditioning on both text and a short acoustic prompt to synthesize high-fidelity speech.
Its extensions—VALL-E X~\cite{zhang2023speak}, VALL-E 2~\cite{chen2024valle2}, and VALL-E R~\cite{han2024valler}—extend the paradigm to cross-lingual synthesis, streaming generation, and improved alignment.
SoCodec~\cite{guo2024socodec} further improves efficiency through semantic-ordered multi-stream tokenization and segment-level modeling.
Through prompt-based conditioning, these models demonstrate strong zero-shot capability and robust speaker generalization.

To reduce inference latency, non-autoregressive (NAR) decoding frameworks have been developed.
SoundStorm~\cite{borsos2023soundstorm} employs a bidirectional Transformer with confidence-based masked token modeling, generating audio tokens in parallel while maintaining autoregressive-level quality.
Multi-token prediction and speculative decoding~\cite{nguyen2025accelerating} further accelerate synthesis by predicting multiple codec tokens per decoding step.
MaskGCT~\cite{wang2024maskgct} adopts a masked generative training strategy inspired by masked language modeling, enabling fast and parallel decoding while supporting in-context learning through prompt-aware input masking.

Building on this foundation, we adapt the MaskGCT framework to singing voice synthesis.
Our approach incorporates structured melody conditioning and improved melody–timbre control in prompt-based synthesis to address the challenges of pitch fidelity and prosody leakage in zero-shot settings.
This design improves controllability over melodic realization while preserving the inference efficiency and generalization strengths of discrete token-based modeling.

\subsection{Prosody and Melody Control in Speech and Singing Voice Synthesis}\label{subsec:prosody_control} 
Fine-grained prosody control—particularly over fundamental frequency (F0) and phoneme duration—is essential for expressive speech and singing synthesis.
Several TTS studies have incorporated explicit prosodic supervision to guide model learning.
Prosody-TTS~\cite{pamisetty2023prosody} augments an end-to-end architecture with auxiliary predictors for phoneme-level F0 and duration, enabling precise rhythm and pitch control without degrading naturalness.
\cite{raitio2022hierarchical} adopt utterance-level prosodic features in a hierarchical non-autoregressive model, providing interpretable style modulation across prosodic dimensions while maintaining synthesis quality.

Recent advances extend prosody control to diffusion-based synthesis.
DiffStyleTTS~\cite{liu2024diffstyletts} combines a diffusion decoder with classifier-free guidance to model prosodic style at both coarse and phoneme-level scales, supporting flexible pitch–duration manipulation.
DrawSpeech~\cite{chen2025drawspeech} enables editing by conditioning on user-drawn pitch–energy sketches, which are refined into high-resolution prosody contours.

In SVS, accurate melody control often requires note-level F0 alignment.
\cite{lee2022expressive} combine dual-path pitch encoders with local style tokens to capture expressiveness beyond score constraints.
Discrete-token SVS frameworks such as TokSing\cite{wu2024toksing} incorporate explicit melody tokens to enrich synthetic singing, while Prompt-Singer~\cite{wang2024prompt} decouples vocal range from melody contour in prompt conditioning to preserve pitch accuracy across timbres.

Despite these advances, few systems address conflicts between external melody guidance and prompt-derived timbre cues in discrete-token SVS.
We address this gap by introducing coarse-to-fine contrastive learning with frame-level pitch supervision to reduce melody leakage from prompts and strengthen external melody control, enabling high-fidelity pitch realization without sacrificing timbre consistency or generalization.

%% file: sections/method.tex
\begin{figure}[t]
\centering
\includegraphics[width=0.9\columnwidth]{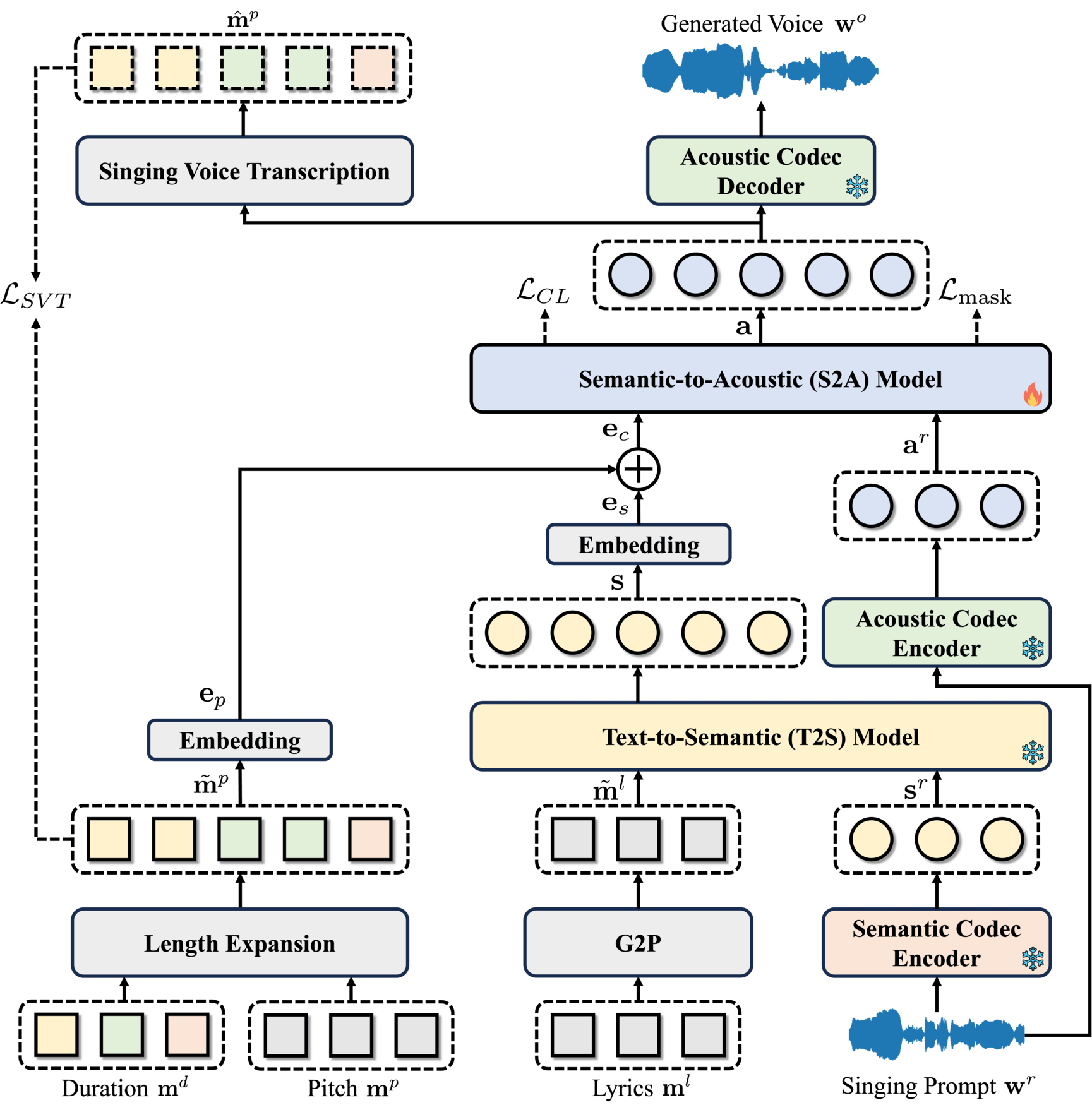} % Reduce the figure size so that it is slightly narrower than the column. Don't use precise values for figure width.This setup will avoid overfull boxes.
\caption{Overview of CoMelSinger. It adopts a two-stage pipeline: a T2S model encodes lyrics into semantic tokens, and an S2A model generates acoustic tokens conditioned on lyrics, pitch, and prompt. SVT provides pitch supervision. All modules except S2A are frozen during training.}
\label{fig:overall}
\end{figure}

\begin{figure*}[t]
  \centering
  \begin{minipage}[t]{0.48\textwidth}
    \centering
    \includegraphics[width=\textwidth]{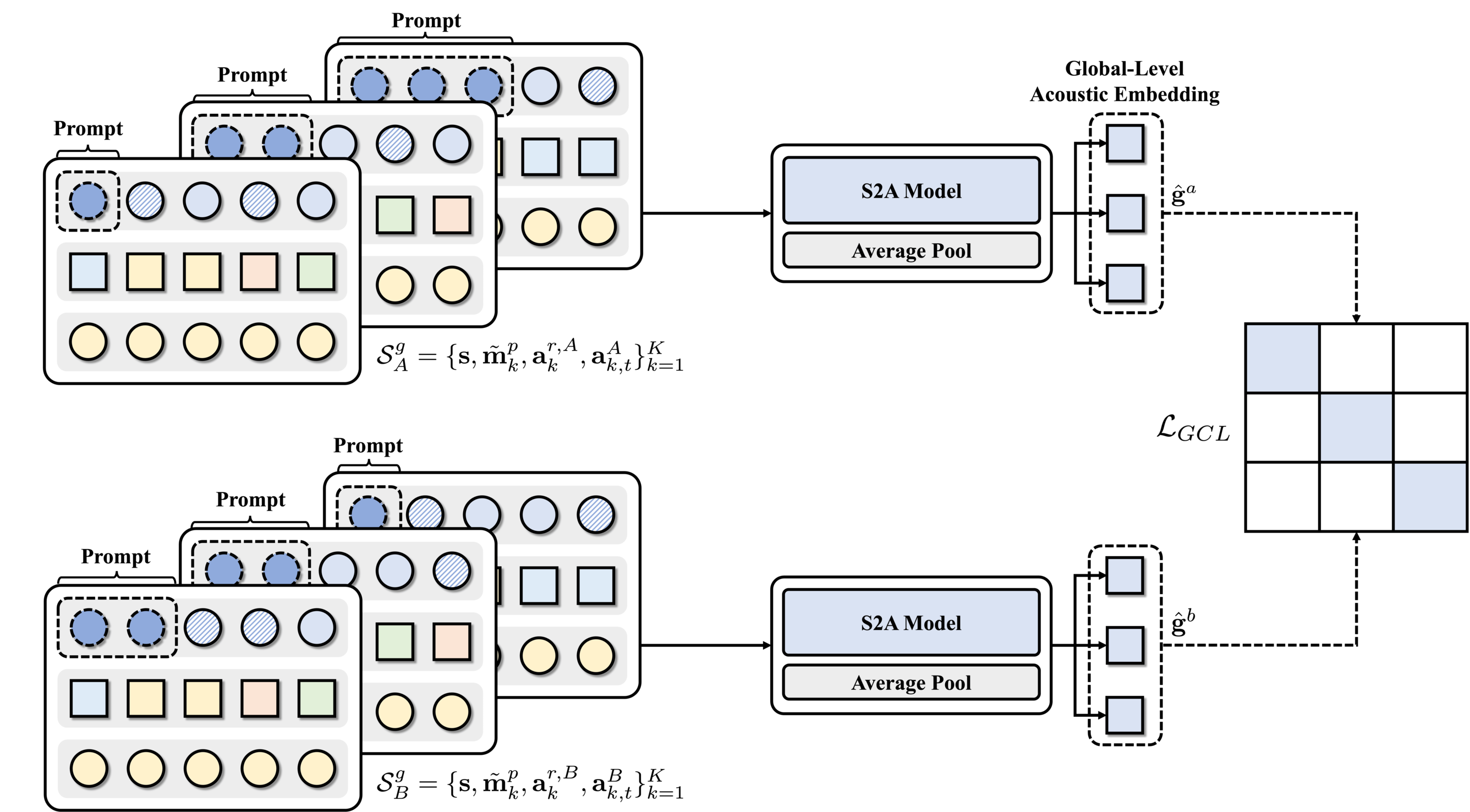}
    (a) Sequence-level Contrastive Learning.
    \label{fig:global}
  \end{minipage}
  \hfill
  \begin{minipage}[t]{0.48\textwidth}
    \centering
    \includegraphics[width=\textwidth]{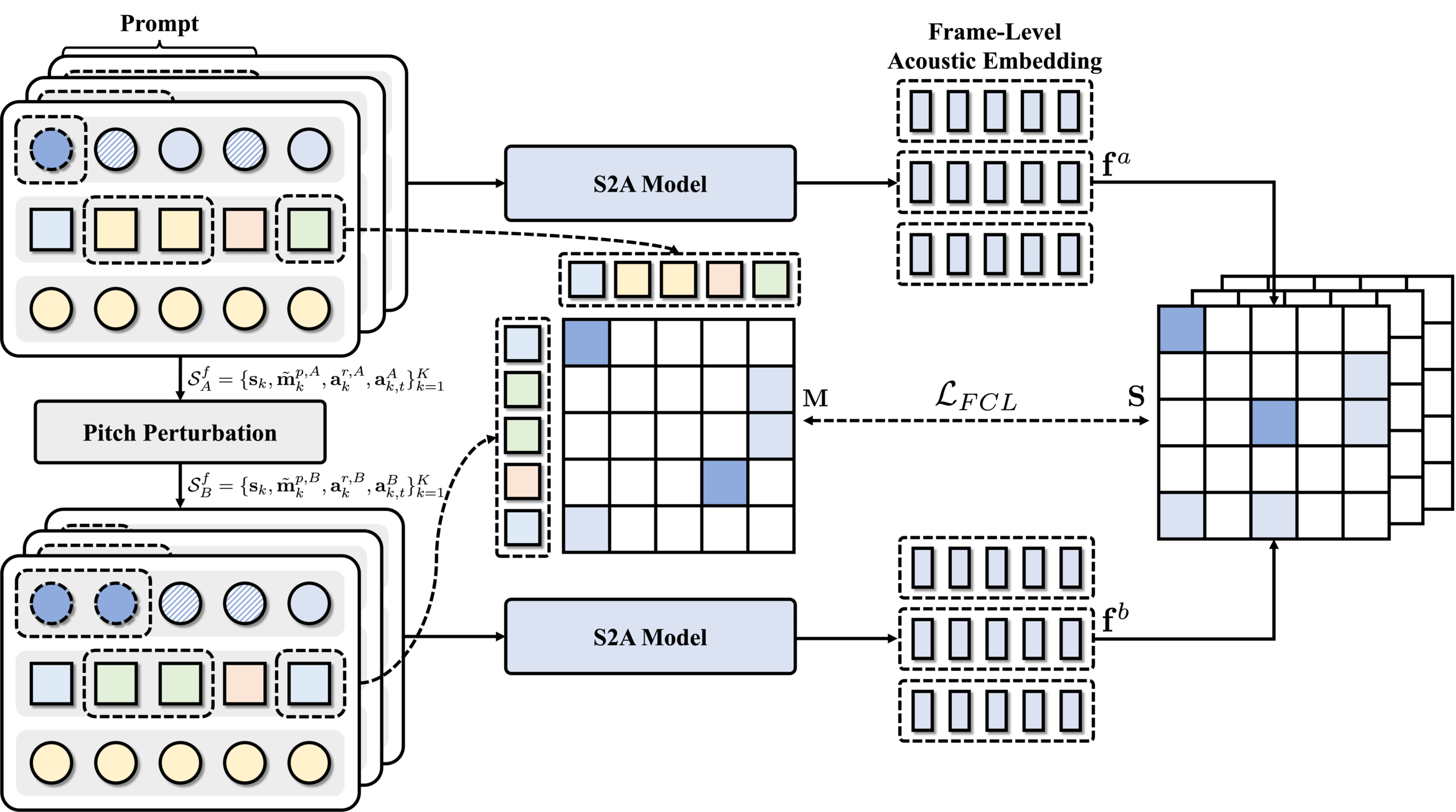}
    (b) Frame-level Contrastive Learning.
    \label{fig:frame}
  \end{minipage}
  \caption{Overview of the coarse-to-fine contrastive learning strategy. (a) Sequence-level contrastive learning encourages timbre consistency across different melodies. (b) Frame-level contrastive learning uses pitch perturbation to enforce local pitch-awareness and disentangle melody from timbre.}
  \label{fig:global_frame}
\end{figure*}

\subsection{Overview}\label{subsec:overview}
To enable zero-shot singing voice synthesis with accurate melody control and disentangled prompt conditioning, we propose \emph{CoMelSinger}, a two-stage framework illustrated in Figure~\ref{fig:overall}. Inspired by MaskGCT~\cite{wang2024maskgct} and Make-A-Voice~\cite{huang2023make}, CoMelSinger comprises a Text-to-Semantic (T2S) stage and a Semantic-to-Acoustic (S2A) stage. The T2S module \(f_{\mathrm{T2S}}\) transforms a lyric token sequence \(\tilde{\mathbf{m}}^{l} = [\tilde{m}^{l}_1, \dots, \tilde{m}^{l}_S] \in \mathcal{V}_{\mathrm{lyr}}^S\), obtained from a Grapheme-to-Phoneme (G2P) converter, and a semantic prompt \(\mathbf{s}^r = E_S(\mathbf{w}^r)\) extracted from the reference waveform \(\mathbf{w}^r\), into a semantic token sequence \(\mathbf{s} \in \mathcal{V}_{\mathrm{sem}}^L\), where \(\mathcal{V}_{\mathrm{sem}}\) denotes the semantic vocabulary and \(L\) the sequence length. The S2A module \(f_{\mathrm{S2A}}\) then predicts acoustic tokens \(\mathbf{a} \in \mathcal{V}_{\mathrm{aco}}^{L \times N}\), conditioned on the semantic tokens \(\mathbf{s}\), an acoustic prompt \(\mathbf{a}^r = E_A(\mathbf{w}^r) \in \mathcal{V}_{\mathrm{aco}}^{L_r \times N}\), and a regulated pitch sequence \(\tilde{\mathbf{m}}^p \in \mathcal{V}_{\mathrm{pit}}^L\). Here, \(N\) denotes the number of residual vector quantization (RVQ) codebooks in the acoustic codec, and the regulated pitch sequence \(\tilde{\mathbf{m}}^p\) is derived from the pitch $\mathbf{m}^p$ and duration $\mathbf{m}^d$ sequence through the length expansion module \(LE(\cdot)\). Both the semantic and acoustic tokens are produced using discrete codec tokenizers following the MaskGCT setup, and the final waveform \(\mathbf{w}^o\) is reconstructed from acoustic tokens via the decoder \(D_A\). The complete pipeline is summarized as:  
\begin{equation}
    \begin{aligned}
        \mathbf{s}^r &= E_S(\mathbf{w}^r), \quad \mathbf{a}^r = E_A(\mathbf{w}^r), \\
        \mathbf{s} &= f_{\mathrm{T2S}}(\mathbf{m}^{l}, \mathbf{s}^r), \\
        \mathbf{a} &= f_{\mathrm{S2A}}(LE(\mathbf{m}^p, \mathbf{m}^d, L), \mathbf{s}, \mathbf{a}^r), \\
        \mathbf{w}^o &= D_A(\mathbf{a}).
    \end{aligned}
\end{equation}

To synchronize pitch information with frame-level features, we map phonetic durations onto the frame index space. Given a pitch sequence $\mathbf{m}^p$ with corresponding durations $\mathbf{m}^d$, the total duration is $D = \sum_i m^d_i$. For each pitch token $m^p_i$, its frame span is derived by rounding the cumulative normalized duration:
\begin{equation}
\begin{aligned}
k_{\text{start},i} = \left\lfloor \tfrac{c_{i-1}}{D} L \right\rceil, \quad
k_{\text{end},i}   = \left\lfloor \tfrac{c_i}{D} L \right\rceil, \quad
c_i = \sum_{j=1}^{i} m^d_j,
\end{aligned}
\end{equation}
where $L$ denotes the length of the semantic feature sequence. The frame-aligned pitch sequence $\tilde{\mathbf{m}}^p$ is then obtained by repeating each $m^p_i$ for $n_i = k_{\text{end}, i} - k_{\text{start}, i}$ frames, ensuring an exact length correspondence with the semantic features.

Following MaskGCT~\cite{wang2024maskgct}, CoMelSinger adopts a non-autoregressive masked generative modeling paradigm for both stages. In this framework, the model learns to reconstruct masked tokens within a sequence conditioned on surrounding context and external inputs, rather than generating tokens sequentially. This allows for parallel decoding and better handling of global context compared to traditional autoregressive models. Specifically, we model the conditional probabilities: 
\(p(\mathbf{s} \mid \mathbf{s}_t; \mathbf{m}^l, \mathbf{s}^r; f_{\theta, \mathrm{T2S}}), \quad
p(\mathbf{a} \mid \mathbf{a}_t; \mathbf{s}, \tilde{\mathbf{m}}^p, \mathbf{a}^r; f_{\theta, \mathrm{S2A}})\). 
Here, \(\mathbf{s}_t\) and \(\mathbf{a}_t\) are the partially masked semantic and acoustic token sequences, and the generation is conditioned on lyric inputs \(\mathbf{m}^l\), pitch-aligned sequence \(\tilde{\mathbf{m}}^p\), and prompt tokens \((\mathbf{s}^r, \mathbf{a}^r)\). 

Building upon the original MaskGCT architecture, which employs a LLaMA-style Transformer backbone \cite{touvron2023llama}, we extend the S2A model by introducing an additional embedding layer for the regulated pitch sequence \(\tilde{\mathbf{m}}^p\). This pitch embedding $\mathbf{e}_p$ is element-wise added to the semantic token embedding  $\mathbf{e}_s$ to form the composite conditioning input $\mathbf{e}_c = \mathbf{e}_p + \mathbf{e}_s$, thereby enabling the model to incorporate both linguistic and melodic information during acoustic token generation.

To further enhance melody controllability and suppress interference from prompt-induced timbre cues, we propose a coarse-to-fine contrastive learning framework. At the sequence level, a contrastive loss encourages the predicted acoustic token sequence \(\mathbf{a}\) to preserve the overall pitch contour defined by \(\tilde{\mathbf{m}}^p\). At the frame level, a token-wise contrastive objective aligns fine-grained acoustic features with localized pitch variations, thereby reinforcing frame-level pitch fidelity. In addition, we introduce an auxiliary singing voice transcription (SVT) model, trained to estimate pitch sequences directly from acoustic tokens. The SVT model provides pseudo pitch labels that serve as external supervision during S2A training. This auxiliary signal further improves alignment between the synthesized melody and the target pitch contour. 

\subsection{Coarse-to-Fine Contrastive Learning}\label{subsec:cl}
Contrastive learning has been increasingly adopted in speech and audio modeling to enforce factor-specific consistency while suppressing undesired variations such as speaker identity or prompt interference. For instance, CLAPSpeech~\cite{ye-etal-2023-clapspeech} applies multi-scale contrastive learning between textual prosody embeddings and corresponding acoustic realizations, improving prosodic expressivity across varying textual contexts. \cite{latif-etal-2021-controlling} propose a contrastive loss to enhance the modeling of prosodic focus—including F0, duration, and intensity—by encouraging TTS systems to distinguish emphasized from neutral phonetic segments. \cite{weston2021learning} use contrastive self-supervision to extract prosody-specific embeddings disentangled from speaker identity, which are useful for style transfer and anonymized generation. \cite{xiao2024contrastive} further explore contrastive pretraining to align textual context with expressive speech realizations, facilitating zero-shot expressive TTS with better generalization. 

Motivated by these findings, our method extends contrastive learning to the discrete-token SVS setting by incorporating hierarchical supervision. At the sequence level, we enforce prompt-invariant acoustic consistency conditioned on identical semantic and pitch tokens, encouraging the model to preserve global melody shape. At the local level, a frame-wise contrastive loss is applied to align acoustic token features with fine-grained pitch variations. This coarse-to-fine scheme allows our model to disentangle pitch conditioning from acoustic prompts and enhances melody fidelity under zero-shot generation.

\subsubsection{Sequence-Level Contrastive Learning}
To promote melody-consistent synthesis under varying acoustic prompts, we introduce a sequence-level symmetric contrastive loss. 
Given a batch of \(K\) training samples that share the same semantic token sequence \(\mathbf{s}\) but differ in pitch sequences \(\tilde{\mathbf{m}}^p_k\), we construct two sets of inputs: 
\(\mathcal{S}^g_A = \{ \mathbf{s}, \tilde{\mathbf{m}}^p_k, \mathbf{a}^{r,A}_k, \mathbf{a}^A_{k, t} \}_{k=1}^{K}\) and
\(\mathcal{S}^g_B = \{ \mathbf{s}, \tilde{\mathbf{m}}^p_k, \mathbf{a}^{r,B}_k, \mathbf{a}^B_{k, t} \}_{k=1}^{K}\),
where $\mathbf{a}^A_{k, t}$ and $\mathbf{a}^B_{k, t}$ denote masked acoustic tokens, \(\mathbf{a}^{r,A}_k\) and \(\mathbf{a}^{r,B}_k\) denote acoustic prompts sampled from distinct utterances of the same singer, ensuring consistent timbre across the pair. 
The prompt length is randomly selected from the range \(\left[ \min\left( \left\lfloor L / 4 \right\rfloor, 5 \right), \left\lfloor L / 2 \right\rfloor \right)\), where \(L\) is the length of the semantic sequence. Each input is processed by the S2A model to produce acoustic token embeddings \(\mathbf{g}^a, \mathbf{g}^b \in \mathbb{R}^{K \times L \times D}\), which are mean-pooled across the time dimension to yield global acoustic representations \(\tilde{\mathbf{g}}^a, \tilde{\mathbf{g}}^b \in \mathbb{R}^{K \times D}\).

To align acoustic outputs with the shared pitch condition while remaining invariant to prompt variation, we apply a symmetric contrastive loss (SCE)~\cite{wang2019symmetric} defined as:
\begin{equation}
\begin{aligned}
\mathcal{L}_{\mathrm{SCL}} & = \frac{1}{2K} \sum_{i=1}^{K} \bigg(
 \log \frac{ \exp\left( \tilde{\mathbf{g}}^a_i \cdot \tilde{\mathbf{g}}^b_i / \tau \right) }
           { \sum_{j=1}^{K} \exp\left( \tilde{\mathbf{g}}^a_i \cdot \tilde{\mathbf{g}}^b_j / \tau \right) } \\
& + \log \frac{ \exp\left( \tilde{\mathbf{g}}^b_i \cdot \tilde{\mathbf{g}}^a_i / \tau \right) }
           { \sum_{j=1}^{K} \exp\left( \tilde{\mathbf{g}}^b_i \cdot \tilde{\mathbf{g}}^a_j / \tau \right) } \bigg),
\end{aligned}
\label{eq:gcl}
\end{equation}
where \(\tau\) is a temperature hyperparameter. Each positive pair \((\tilde{\mathbf{g}}^{a}_i, \tilde{\mathbf{g}}^{b}_i)\) corresponds to index-aligned embeddings generated under identical semantic tokens \(\mathbf{s}\) and regulated pitch sequence \(\tilde{\mathbf{m}}^p\), but conditioned on different acoustic prompts \(\mathbf{a}^{r,A}_i\) and \(\mathbf{a}^{r,B}_i\). These prompts are randomly selected from non-overlapping segments of the same singer's recordings, ensuring consistent timbre while introducing natural variation. In contrast, off-diagonal pairs \((\tilde{\mathbf{g}}^{a}_i, \tilde{\mathbf{g}}^{b}_j)\) for \(i \ne j\) serve as negatives due to mismatched pitch sequences, even though semantic tokens and speaker identity remain the same. These negatives are nontrivial, as they reflect realistic melodic differences under otherwise comparable contextual and timbral conditions. This design encourages the model to focus on capturing global pitch structure while remaining invariant to prompt-induced variability. 

While this sequence-level objective enforces high-level melodic consistency, it does not explicitly supervise token-level alignment. We therefore complement it with a frame-level contrastive loss to further enhance fine-grained melody control.

\subsubsection{Frame-Level Contrastive Learning}\label{subsec:framecl}

While the sequence contrastive loss promotes utterance-level melody consistency, it does not directly enforce fine-grained pitch alignment at the frame level—an essential factor in singing synthesis due to rapid and expressive melodic changes. To address this limitation, we introduce a frame-level contrastive objective that supervises token-wise alignment between generated acoustic representations and the input pitch contour. Our design is motivated by recent work such as CTAP~\cite{qiang2024learning}, which leverage contrastive learning to align discrete phoneme sequences with speech features for TTS, voice conversion, and ASR tasks under limited supervision. Although our formulation differs in both granularity and modality—operating on pitch tokens rather than phonemes, and targeting melody alignment in singing synthesis—these studies underscore the effectiveness of contrastive supervision for bridging symbolic and acoustic representations. By extending this idea to the SVS domain, our frame-level contrastive loss enhances local pitch fidelity while remaining robust to prompt-induced variation, thereby enabling more precise and expressive melody control in zero-shot scenarios.

Given a batch of \(K\) training samples with distinct semantic token sequences \(\mathbf{s}_k\) and corresponding pitch sequences \(\mathbf{m}^p_k\), we construct two sets of inputs:
\(
\mathcal{S}^f_A = \{ \mathbf{s}_k, \tilde{\mathbf{m}}^{p,A}_k, \mathbf{a}^{r,A}_k, \mathbf{a}^A_{k,t} \}_{k=1}^{K}\) and \(\mathcal{S}^f_B = \{ \mathbf{s}_k, \tilde{\mathbf{m}}^{p,B}_k, \mathbf{a}^{r,B}_k, \mathbf{a}^B_{k,t} \}_{k=1}^{K}, 
\)
where $\mathbf{a}^{A}_{k, t}$ and $\mathbf{a}^{B}_{k, t}$ denote the masked acoustic tokens, \(\tilde{\mathbf{m}}^{p,A}_k\) and \(\tilde{\mathbf{m}}^{p,B}_k\) denote the regulated pitch sequences derived from the original pitch tokens \(\mathbf{m}^p_k\) and their perturbed variants \(P(\mathbf{m}^p_k)\), respectively. The acoustic prompts \(\mathbf{a}^{r,A}_k\) and \(\mathbf{a}^{r,B}_k\) are sampled from different utterances of the same singer to ensure consistent timbre across pairs. 
The perturbation function \(P(\cdot)\) offsets 50\% of pitch tokens by integers randomly sampled from \([-6, 6]\), while preserving the original duration sequence \(\mathbf{m}^d\).

Each input is passed through the S2A model to produce frame-level acoustic embeddings \(\mathbf{f}^a, \mathbf{f}^b \in \mathbb{R}^{K \times L \times D}\). For each training sample \(k\), we compute a cosine similarity matrix \(\mathbf{S}^k \in \mathbb{R}^{L \times L}\) between \(\mathbf{f}^a\) and \(\mathbf{f}^b\). To supervise the similarity learning, we define a soft label matrix \(\mathbf{Y}^k \in [-1, 1]^{L \times L}\) capturing pitch and semantic alignment:
\begin{equation}
Y^k_{ij} =
\begin{cases}
1, & \text{if } \tilde{m}^{p,A}_{k,i} = \tilde{m}^{p,B}_{k,j} \land s_i = s_j \\
\alpha, & \text{if } \tilde{m}^{p,A}_{k,i} = \tilde{m}^{p,B}_{k,j} \land s_i \ne s_j \\
0, & \text{if } \tilde{m}^{p,A}_{k,i} \ne \tilde{m}^{p,B}_{k,j} \\
-1, & \text{if either frame is silence or padding}
\end{cases},
\label{eq:framecl_labels}
\end{equation}
where \(\alpha \in (0, 1)\) is a tunable coefficient that softly downweights semantically mismatched but pitch-aligned pairs. The frame-level contrastive loss is formulated as a masked regression objective:
\begin{equation}
\mathcal{L}_{\text{FCL}} = \frac{1}{K} \sum_{k=1}^{K} \sum_{i=1}^{L} \sum_{j=1}^{L} 
\mathds{1}\left[ Y^k_{ij} \geq 0 \right] \cdot \left( S^k_{ij} - Y^k_{ij} \right)^2,
\label{eq:fcl}
\end{equation}
where \(\mathds{1}[\cdot]\) is an indicator function that masks out invalid entries (e.g., silence or padding).

This formulation encourages the model to produce highly similar acoustic embeddings when both pitch and semantic content align, moderately similar embeddings when only pitch aligns, and dissimilar embeddings otherwise. The similarity is computed within each utterance because the vocal range is typically locally bounded, making repeated pitch tokens more likely. In contrast, pitch overlap across utterances is rare and thus excluded. Additionally, even within the same utterance, repeated pitches may be associated with different semantic tokens, resulting in subtle acoustic variation. Our soft labeling mechanism accounts for this by assigning intermediate similarity, thereby avoiding over-penalization while promoting melody-consistent synthesis. Hence, the final contrastive learning objective is as follows: 
\begin{equation}
    \mathcal{L}_{\text{CL}} = \lambda_{\text{SCL}} \cdot \mathcal{L}_{\text{SCL}} + \lambda_{\text{FCL}} \cdot \mathcal{L}_{\text{FCL}},
    \label{eq:cl}
\end{equation}
\noindent
where \(\lambda_{\text{SCL}}\) and \(\lambda_{\text{FCL}}\) are weighting coefficients that balance the contributions of sequence-level and frame-level supervision, respectively.

\subsection{Singing Voice Transcription for Pitch Guidance}\label{subsec:SVT_pg} 
Recent studies have explored the use of Singing Voice Transcription (SVT) to support singing voice synthesis (SVS). For example, ROSVOT~\cite{li2024robust} proposes a robust SVT model to produce high-quality pitch annotations for large-scale singing datasets, thereby improving SVS performance by enhancing training data quality. However, such approaches treat SVT as an independent preprocessing tool, disconnected from the synthesis process.

In contrast, we integrate the SVT module directly into the training pipeline to provide explicit frame-level pitch supervision, thereby enhancing melody modeling and alignment. Specifically, the SVT model predicts frame-wise discrete pitch tokens from acoustic codec representations, which are then compared against the ground-truth pitch sequence. This supervision enforces alignment between the generated acoustic tokens and the intended melody, encouraging consistent pitch realization, particularly under zero-shot conditions. Furthermore, as the SVT module operates entirely on discrete representations, it is naturally compatible with our codec-based SVS framework and does not require raw audio or continuous F0 contours.

The SVT model adopts a lightweight encoder-only Transformer architecture. Given a sequence of acoustic tokens $\mathbf{a}$, it predicts the corresponding pitch token sequence $\hat{\mathbf{m}}^p$ of length $L$. The encoder consists of four Transformer layers with a hidden size of 512 and eight attention heads. Each input frame comprises 12 discrete acoustic codes, which are individually embedded, concatenated, and projected to a 512-dimensional representation, followed by layer normalization. The resulting embedding sequence is then passed through a linear classification head to predict a pitch token for each frame. The model is trained using a standard cross-entropy loss between the predicted and reference pitch sequences.

To provide frame-level supervision for pitch modeling, we leverage the pretrained SVT model as a pitch predictor to generate pseudo labels in the form of pitch token sequences \(\tilde{\mathbf{m}}^p\), which are temporally aligned with the acoustic frames. As the primary training objective, we apply a cross-entropy loss \(\mathcal{L}_{\rm CE}\) between the predicted pitch tokens \(\hat{\mathbf{m}}^p\) and the SVT-derived ground-truth \(\tilde{\mathbf{m}}^p\), encouraging accurate token-level classification. However, the cross-entropy objective alone does not account for the temporal continuity inherent in repeated pitch tokens. This often results in jittery predictions, fragmented note segments, and rhythmically unstable outputs. 

\subsubsection{Segment Transition Loss}
Prior studies have highlighted the importance of modeling temporal structure for natural singing synthesis. XiaoiceSing~\cite{lu2020xiaoicesing} introduces syllable-level duration modeling to preserve rhythmic consistency, while Singing-Tacotron~\cite{wang2022singing} enhances segmental alignment through transition tokens and duration-informed attention mechanisms.

Motivated by these findings, we propose a segment transition loss \(\mathcal{L}_{\text{seg}}\) to impose structural regularity on the predicted pitch token sequence. Let \(\mathbf{p} \in \mathbb{R}^{L \times C}\) denote the predicted frame-level pitch token distribution obtained by applying a softmax to the decoder logits, where \(L\) is the number of frames and \(C\) is the size of the pitch token vocabulary. The loss is defined as follows:
\begin{equation}
\begin{aligned}
\mathcal{L}_{\text{seg}} 
&= \sum_{t=2}^{L} 
\Big[
(1 - b_t) \cdot \|\mathbf{p}_t - \mathbf{p}_{t-1}\|^2 \\
&\quad + b_t \cdot \max\!\left(0, \delta - \|\mathbf{p}_t - \mathbf{p}_{t-1}\|\right)^2
\Big], 
\end{aligned}
\end{equation}
where \(b_t = \mathds{1}[\tilde{m}^p_t \ne \tilde{m}^p_{t-1}]\) is a binary indicator marking ground-truth pitch boundaries, and \(\delta\) is a fixed margin that enforces dissimilarity across transitions. This formulation penalizes minimal variation within sustained pitch regions while promoting sharper contrast at pitch change boundaries, thereby enhancing segment continuity and expressive phrasing in the synthesized output. 

\subsubsection{Soft Duration Loss}
Inspired by recent advances in speech synthesis that emphasize the importance of temporal alignment and duration modeling~\cite{ren2019fastspeech, lu2020xiaoicesing}, we introduce a soft duration loss \(\mathcal{L}_{\text{dur}}\) to enhance the rhythmic fidelity of frame-level pitch predictions. Prior works such as FastSpeech~\cite{ren2019fastspeech} and XiaoiceSing~\cite{lu2020xiaoicesing} employ explicit duration predictors or auxiliary alignment modules to supervise temporal structures. While effective, these methods often introduce architectural overhead or struggle to generalize in expressive singing scenarios. In contrast, our approach provides a fully differentiable supervision signal by directly supervising the temporal distribution of pitch token probabilities using the softmax outputs of the model, without requiring any external duration modeling.

Given a symbolic duration sequence \(\mathbf{m}^d = [m_1^d, \ldots, m_S^d]\), we normalize it into a frame-level allocation \(\mathbf{a}^d = [a_1^d, \ldots, a_S^d]\), where each element is computed as \(a_i^d = \left\lfloor \frac{m_i^d}{D} \cdot L \right\rfloor\), where \(D = \sum\limits_{i=1}^{S} m^d_i\) and \(L\) denotes the total number of frames. Let \(\mathbf{p} \in \mathbb{R}^{L \times C}\) denote the predicted frame-level pitch token distribution obtained via softmax, where \(C\) is the size of the pitch vocabulary. For each target pitch token \(m_i^p\), we define its soft duration as the cumulative probability mass \(\mathbf{p}_t[m_i^p]\) over its allocated segment of length \(a_i^d\). The soft duration loss is given by
\begin{equation}
\mathcal{L}_{\text{dur}} = \sum\limits_{i=1}^{S} \left( \sum\limits_{t = T_i}^{T_i + a_i^d - 1} \mathbf{p}_t[m_i^p] - a_i^d \right)^2,
\end{equation}
where \(T_i = \sum\limits_{j=1}^{i-1} a_j^d\) denotes the starting frame index for the \(i\)th pitch token.

This formulation encourages the model to allocate appropriate probability mass to each pitch token across time, thereby promoting temporally coherent and rhythmically faithful melody generation. The final training objective for the SVT module combines the cross-entropy loss, segment transition loss, and soft duration loss:
\begin{equation}
\mathcal{L}_{\text{SVT}} = \mathcal{L}_{\text{CE}} + \lambda_{\text{seg}} \cdot \mathcal{L}_{\text{seg}} + \lambda_{\text{dur}} \cdot \mathcal{L}_{\text{dur}}.
\label{eq:svt}
\end{equation}

\subsection{Training and Inference Procedures}\label{subsec:train_inf} 

\begin{algorithm}[t]
\caption{Finetuning S2A with Contrastive Learning and SVT Supervision}
\label{alg:training}
\begin{algorithmic}[1]
\REQUIRE S2A model parameters $\theta$; frozen SVT model $f_{\text{SVT}}$; training set $\mathcal{D}$; loss weights $\lambda_{\text{CL}}, \lambda_{\text{SVT}}, \lambda_{\text{mask}}$; number of epochs $N$; learning rate $\eta$
\ENSURE Trained parameters $\theta$
\FOR{$i = 1$ to $N$}
    \STATE Sample batch $\mathcal{B} = \{ \mathbf{x}_1, \dots, \mathbf{x}_K \}$ from $\mathcal{D}$
    \STATE \textbf{Split} $\mathcal{B}$ into $\mathcal{S}^g_A = \{ \mathbf{x}_1, \dots, \mathbf{x}_{K_g} \}$ and $\mathcal{S}^f_A = \{ \mathbf{x}_{K_g+1}, \dots, \mathbf{x}_K \}$
    \STATE \textbf{Construct} $\mathcal{S}^g_B = \mathcal{S}^g_A$, $\mathcal{S}^f_B \gets P(\mathcal{S}^f_A)$
    \STATE Define $\mathcal{B}' \gets \mathcal{S}^g_B \cup \mathcal{S}^f_B$
    \STATE $\mathbf{a}^r \gets \texttt{PromptGen}(\mathcal{B})$, $\mathbf{a}^{r'} \gets \texttt{PromptGen}(\mathcal{B}')$
    \STATE $\tilde{\mathcal{B}} \gets \mathcal{B} \cup \mathbf{a}^r$, $\tilde{\mathcal{B}}' \gets \mathcal{B}' \cup  \mathbf{a}^{r'}$
    \STATE \textbf{Forward pass:}
    \STATE \quad $\mathbf{e} \gets \hat{f}_{\text{S2A}}(\tilde{\mathcal{B}})$, $\mathbf{e}' \gets \hat{f}_{\text{S2A}}(\tilde{\mathcal{B}}')$
    \STATE \quad Split $\mathbf{e}$ into $\mathbf{g}^a = \mathbf{e}_{1:K_g}$ and  $\mathbf{f}^a = \mathbf{e}_{K_g+1:K}$
    \STATE \quad Split $\mathbf{e}'$ into $\mathbf{g}^b = \mathbf{e}'_{1:K_g}$ and $\mathbf{f}^b = \mathbf{e}'_{K_g+1:K}$
    \STATE \quad $\hat{\mathbf{g}}^a \gets \texttt{AvgPool}(\mathbf{g}^a)$,  $\hat{\mathbf{g}}^b \gets \texttt{AvgPool}(\mathbf{g}^b)$

    \STATE \textbf{Contrastive losses:}
    \STATE \quad Compute $\mathcal{L}_{\text{SCL}}$ from $\hat{\mathbf{g}}^a$ and $\hat{\mathbf{g}}^b$ according to~\eqref{eq:gcl}
    \STATE \quad Compute $\mathcal{L}_{\text{FCL}}$ from $\mathbf{f}^a$ and $\mathbf{f}^b$ according to~\eqref{eq:fcl} 
    \STATE \quad Compute $\mathcal{L}_{\text{CL}}$ according to \eqref{eq:cl}
    \STATE \textbf{Mask prediction and SVT loss:}
    \STATE \quad $\hat{\mathbf{a}} \gets f_{\text{head}}(\mathbf{f}^a)$
    \STATE \quad $\mathcal{L}_{\text{mask}} \gets \texttt{MaskLoss}(\hat{\mathbf{a}}, \mathbf{a})$
    \STATE \quad $\hat{\mathbf{m}}^p \gets f_{\text{SVT}}(\texttt{StopGrad}(\hat{\mathbf{a}}))$
    \STATE \quad Compute $\mathcal{L}_{\text{SVT}}$ from $\hat{\mathbf{m}}^p$ and $\tilde{\mathbf{m}}^p$ according to~\eqref{eq:svt}
    \STATE \textbf{Total loss and update:}
    \STATE \quad $\mathcal{L} \gets \lambda_{\text{CL}} \cdot \mathcal{L}_{\text{CL}} + \lambda_{\text{SVT}} \cdot \mathcal{L}_{\text{SVT}} + \lambda_{\text{mask}} \cdot \mathcal{L}_{\text{mask}}$
    \STATE \quad $\theta \gets \theta - \eta \cdot \nabla_{\theta} \mathcal{L}$
\ENDFOR
\end{algorithmic}
\end{algorithm}

We begin by training the Singing Voice Transcription (SVT) model to provide frame-level pitch supervision for subsequent Semantic-to-Acoustic (S2A) adaptation. The SVT model is optimized using a cross-entropy loss $\mathcal{L}_{\text{CE}}$ between the regulated pitch token sequence $\tilde{\mathbf{m}}^p$ and the acoustic token sequence $\mathbf{a}$, thereby learning to predict temporally aligned pitch trajectories from acoustic inputs. Once trained, the SVT module is frozen to serve as a fixed auxiliary supervisor during S2A training.

We then fine-tune the S2A model built upon the MaskGCT framework~\cite{wang2024maskgct}, which leverages masked acoustic modeling for non-autoregressive generation. The training objective for the S2A model comprises three components: (1) the mask token prediction loss $\mathcal{L}_{\text{mask}}$, which reconstructs randomly masked acoustic tokens from noisy inputs using a masked denoising objective; (2) a coarse-to-fine contrastive loss $\mathcal{L}_{\text{CL}}$, composed of sequence-level ($\mathcal{L}_{\text{SCL}}$) and frame-level ($\mathcal{L}_{\text{FCL}}$) terms, to enforce consistency between the melody condition and generated acoustic tokens while mitigating prosody leakage from the prompt; and (3) an auxiliary SVT loss $\mathcal{L}_{\text{SVT}}$, which encourages the predicted acoustic tokens to be rhythmically and melodically consistent with the SVT-inferred pitch contour. The fine-tuning algorithm for the S2A model is summarized in Algorithm~\ref{alg:training}. Each training sample \(\mathbf{x}_k \in \mathcal{B}\) comprises a semantic sequence \(\mathbf{s}_k\), a regulated pitch sequence \(\tilde{\mathbf{m}}^p_k\), and a time-aligned acoustic token sequence \(\mathbf{a}_{k,t}\). 

To improve adaptation efficiency and reduce overfitting, we apply Low-Rank Adaptation (LoRA)~\cite{hu2022lora} to fine-tune the S2A model's diffusion estimator module, which is implemented using a DiffLlama-style architecture. LoRA introduces trainable low-rank matrices into the linear layers of pretrained models, enabling efficient fine-tuning by updating only a small subset of parameters while keeping the original weights frozen. This allows our model to retain prior knowledge from the TTS domain while adapting to the stylistic nuances of singing voice synthesis with limited data.

During inference, we follow the parallel iterative decoding introduced in MaskGCT~\cite{wang2024maskgct} to generate acoustic token sequences. Unlike the original setup, which requires either an explicit duration input or a learned duration predictor to determine the output length, we directly use the ground-truth duration sequence $\mathbf{m}^d$ extracted from the input music score. This allows precise control over the number of generated tokens---equal to the total duration $D = \sum\limits_i m_i^d$---thus preserving the intended temporal structure of the synthesized performance.

%% file: sections/exp.tex
\subsection{Dataset}
We conduct experiments on two publicly available mandarin singing corpora: M4Singer \cite{zhang2022m4singer}\footnote{\url{https://github.com/M4Singer/M4Singer}} and Opencpop \cite{wang2022opencpop}\footnote{\url{https://xinshengwang.github.io/opencpop/}}. The M4Singer dataset comprises studio-quality recordings from 20 professional singers spanning SATB vocal ranges, along with comprehensive annotations including lyrics, pitch, note duration, and slur information. The Opencpop dataset contains 100 Chinese pop songs sung by a professional female vocalist, with precise phoneme, note, and syllable-level annotations aligned to the music score. For evaluation, we construct both seen- and unseen-singer test sets. For seen-singer evaluation, we randomly select 50 utterances each from the M4Singer and Opencpop datasets. For zero-shot (unseen-singer) evaluation, we use 10 male and 10 female singers from the OpenSinger \cite{huang2021multi}\footnote{\url{https://github.com/Multi-Singer/Multi-Singer.github.io?tab=readme-ov-file}} dataset. Since OpenSinger lacks complete music score annotations, we pair it with M4Singer’s score sequences to enable evaluation. All audio is uniformly down-sampled to 24 kHz with 16-bit quantization. 

To fine-tune the S2A model, we preprocess the dataset to obtain temporally aligned semantic tokens \(\mathbf{s}\), acoustic tokens \(\mathbf{a}\), and regulated pitch tokens \(\tilde{\mathbf{m}}^p\). All audio segments are first converted to mono and resampled to 24~kHz. We then utilize the pretrained semantic and acoustic codec models from the MaskGCT framework\footnote{\url{https://github.com/open-mmlab/Amphion/tree/main/models/tts/maskgct}} to extract \(\mathbf{s}\) and \(\mathbf{a}\). For samples with mismatched token lengths, we apply zero-padding to align their temporal dimensions for frame-level supervision. The regulated pitch sequence \(\tilde{\mathbf{m}}^p\) is derived by expanding the original pitch sequence \(\mathbf{m}^p\) according to the duration sequence \(\mathbf{m}^d\), as described in Section~\ref{subsec:overview}. The SVT model is trained using the preprocessed acoustic tokens \(\mathbf{a}\) and the corresponding regulated pitch tokens \(\tilde{\mathbf{m}}^p\).

\subsection{Implementation Details}
The SVT model is trained on a single NVIDIA RTX A5000 GPU using the AdamW optimizer with a learning rate of 1e\textminus5, weight decay of 0.01, and a cosine learning rate schedule with 5K warm-up steps over 50K updates. Training is performed for 100 epochs with a batch size of 32 using mixed-precision (FP16) computation. Subsequently, the S2A model is fine-tuned on four NVIDIA RTX A5000 GPUs with data parallelism. We adopt the AdamW optimizer with a learning rate of 1e\textminus5, 32K warm-up steps, and the inverse square root learning rate schedule. Fine-tuning is conducted for 300K steps with a total batch size of 32, where the first 8 samples are used for sequence contrastive learning and the remaining 24 for frame-level contrastive learning. We apply dropout (0.1), label smoothing (0.1), and gradient clipping to stabilize training. During this stage, all model components are frozen except the S2A decoder. The loss weights are set as follows: \(\lambda_{\text{SCL}} = 0.5\), \(\lambda_{\text{FCL}} = 1.0\), \(\lambda_{\text{CL}} = 0.1\), \(\lambda_{\text{seg}} = 0.5\), \(\lambda_{\text{dur}} = 0.3\), and \(\lambda_{\text{SVT}} = 0.2\).

\subsection{Evaluation Metrics}
% \textit{To evaluate our Singing Voice Transcription (SVT) model, ...} 
\subsubsection{Objective Evaluation}
To quantify the performance of our system in terms of pitch accuracy, timbre consistency, and perceptual quality, we conduct objective evaluations under both seen-singer and zero-shot settings. The following metrics are employed:

\paragraph{Mel-Cepstral Distortion (MCD)}  
MCD is used to evaluate spectral fidelity by computing the frame-wise Euclidean distance between mel-cepstral coefficients of the synthesized and reference audio. It serves as a proxy for spectral similarity, where lower values indicate more accurate spectral reconstruction and reduced distortion.

\paragraph{Fundamental Frequency RMSE (F0-RMSE)}  
F0-RMSE measures pitch prediction accuracy by calculating the root mean squared error between the fundamental frequency (F0) trajectories of the generated and reference waveforms. A lower F0-RMSE reflects better alignment with the intended melody and more precise pitch control.

\paragraph{Speaker Embedding Cosine Similarity (SECS)} 
To assess timbre similarity, we compute cosine similarity between speaker embeddings extracted from the synthesized and reference audio using a WavLM-based speaker verification model~\cite{chen2022wavlm}\footnote{\url{https://huggingface.co/microsoft/wavlm-base-sv}}. SECS values range from 0 to 1, with higher scores indicating closer alignment in vocal identity.

\paragraph{SingMOS}  
SingMOS (Singing Mean Opinion Score)~\cite{tang2024singmos}\footnote{\url{https://github.com/South-Twilight/SingMOS}} is a learned metric trained to predict human perceptual ratings of singing voice quality. It is based on a curated dataset of professional listening tests, in which human raters assign mean opinion scores to both natural and synthesized singing in Chinese and Japanese, addressing the scarcity of large-scale perceptual annotations in the singing domain. SingMOS produces scores in the range of 0 to 5, with higher values indicating greater perceived naturalness and overall quality. As a reference-free metric, it enables scalable automatic evaluation in zero-shot and low-resource conditions. 

\subsubsection{Subjective Evaluation}
To assess perceptual quality, we conducted a Mean Opinion Score (MOS) evaluation with 20 participants who have formal training in singing and experience in vocal performance\footnote{This study has been approved by the Department Ethics Review Committee
(DERC) at the National University of Singapore under 
DERC Ref Code: 000479.}. Each participant rated the synthesized samples based on overall naturalness (MOS-N), audio quality (MOS-Q), and timbre similarity (SMOS). A 5-point Likert scale was used, where a score of 5 indicates excellent perceptual quality and 1 denotes poor quality. 

%% file: sections/result.tex
% \subsection{Main Results}

\subsection{Evaluating Prompt-Induced Prosody Similarity in MaskGCT}

\begin{table*}[t]
\centering
\caption{The prosody similarity between synthesized and prompt speech in terms of differences in pitch, energy, and other prosodic indicators. Lower values indicate higher similarity.}
\label{tab:prosody-leakage}
% \resizebox{\textwidth}{!}{
\begin{tabular}{lcccc|cccc|ccc}
\toprule
\multirow{2}{*}{\textbf{LibriTTS}} 
& \multicolumn{4}{c}{\textbf{Pitch}} 
& \multicolumn{4}{c}{\textbf{Energy}} 
& \multicolumn{3}{c}{\textbf{Others}} \\
\cmidrule(lr){2-5} \cmidrule(lr){6-9} \cmidrule(lr){10-12}
& Mean $\downarrow$ & Std $\downarrow$ & Skew $\downarrow$ & Kurt $\downarrow$
& Mean $\downarrow$ & Std $\downarrow$ & Skew $\downarrow$ & Kurt $\downarrow$
& Jitter $\downarrow$ & Shimmer $\downarrow$ & HNR $\downarrow$ \\
\midrule
\rowcolor{gray!10}
Paired   & \textbf{19.03} & \textbf{22.79} & \textbf{2.43} & \textbf{19.68} & \textbf{1.80} & \textbf{1.54} & \textbf{0.33} & \textbf{0.95} & \textbf{0.63} & \textbf{0.56} & \textbf{3.28} \\
Unpaired & 53.41 & 32.47 & 2.81 & 21.82 & 4.60 & 2.28 & 0.46 & 1.23 & 0.89 & 0.78 & 3.79 \\
\midrule
\midrule
\multirow{2}{*}{\textbf{AISHELL-3}} 
& \multicolumn{4}{c}{\textbf{Pitch}} 
& \multicolumn{4}{c}{\textbf{Energy}} 
& \multicolumn{3}{c}{\textbf{Others}} \\
\cmidrule(lr){2-5} \cmidrule(lr){6-9} \cmidrule(lr){10-12}
& Mean $\downarrow$ & Std $\downarrow$ & Skew $\downarrow$ & Kurt $\downarrow$
& Mean $\downarrow$ & Std $\downarrow$ & Skew $\downarrow$ & Kurt $\downarrow$
& Jitter $\downarrow$ & Shimmer $\downarrow$ & HNR $\downarrow$ \\
\midrule
\rowcolor{gray!10}
Paired   & \textbf{42.70} & \textbf{26.29} & \textbf{1.13} & \textbf{3.85} & \textbf{2.94} & \textbf{2.38} & \textbf{0.44} & \textbf{1.71} & \textbf{0.78} & \textbf{0.43} & \textbf{4.42} \\
Unpaired & 65.73 & 29.87 & 1.63 & 7.05 & 5.05 & 2.58 & 0.57 & 1.83 & 0.94 & 0.58 & 4.95 \\
\bottomrule
\end{tabular}
% }
\end{table*}

\begin{table*}[t]
\centering
\caption{Evaluation results on the seen test set for singing voice synthesis. Subjective metrics are reported with 95\% confidence intervals. GT stands for Ground Truth.}
\label{tab:main-eval-seen}
% \resizebox{\textwidth}{!}{
\begin{tabular}{lccc|cccc}
\toprule
\multirow{2}{*}{Model} 
& \multicolumn{3}{c}{\textbf{Subjective Evaluations}} 
& \multicolumn{4}{c}{\textbf{Objective Evaluations}} \\
\cmidrule(lr){2-4} \cmidrule(lr){5-8}
& MOS-Q $\uparrow$ & MOS-N $\uparrow$ & SMOS $\uparrow$
& MCD $\downarrow$ & F0-RMSE $\downarrow$ 
& SingMOS $\uparrow$ & SECS $\uparrow$ \\
\midrule
GT                      & 4.17 $\pm$ 0.16 & 4.38 $\pm$ 0.18 & 4.41 $\pm$ 0.14 & -    & -     & 4.37 & 0.925 \\
GT (Acoustic Codec)              & 4.01 $\pm$ 0.22 & 4.19 $\pm$ 0.19 & 4.48 $\pm$ 0.12 & 0.93 & 0.012 & 4.31 & 0.906 \\
\midrule
DiffSinger~\cite{liu2022diffsinger}         & 3.68 $\pm$ 0.20 & 3.79 $\pm$ 0.15 & 3.86 $\pm$ 0.16 & 4.59 & 0.084 & 4.13 & 0.769 \\
VISinger2~\cite{zhang2022visinger2}         & 3.59 $\pm$ 0.22 & 3.86 $\pm$ 0.16 & 3.91 $\pm$ 0.16 & 5.36 & 0.061 & 4.15 & 0.792 \\
StyleSinger~\cite{zhang2024stylesinger}  & 3.67 $\pm$ 0.15 & 3.92 $\pm$ 0.21 & 4.11 $\pm$ 0.16 & 4.95 & 0.112 & 4.19 & 0.833 \\
SPSinger~\cite{zhao2025spsinger}       & 3.81 $\pm$ 0.18 & \textbf{4.10 $\pm$ 0.12} & 4.06 $\pm$ 0.17 & 4.28 & 0.054 & 4.28 & 0.860 \\
Vevo 1.5\cite{zhang2025vevo2} & 3.85 $\pm$ 0.12 & 3.96 $\pm$ 0.16 & 4.17 $\pm$ 0.16 & 4.18 & 0.051 & \textbf{4.39} & 0.907 \\
\midrule
\rowcolor{gray!10}
CoMelSinger (ours)    & \textbf{3.90 $\pm$ 0.16} & 4.02 $\pm$ 0.12 & \textbf{4.22 $\pm$ 0.15} 
& \textbf{4.17} & \textbf{0.042} &4.32 & \textbf{0.912} \\
\bottomrule
\end{tabular}
% }
\end{table*}

Several recent TTS systems~\cite{shen2023naturalspeech, ju2024naturalspeech, li-etal-2025-styletts} have reported high prosodic similarity between the speech prompt and the synthesized output. While this may appear beneficial in TTS, it reveals a form of prosody leakage, where expressive cues from the prompt inadvertently influence the generated speech. This issue becomes particularly problematic in singing voice synthesis (SVS), where pitch and rhythm should be governed solely by the input music score. As discussed in Section~\ref{sec:intro}, we refer to this phenomenon as prosody leakage. 

To investigate whether MaskGCT~\cite{wang2024maskgct} exhibits such behavior, we conduct a prosody similarity analysis following prior evaluation protocols. NaturalSpeech 2 and 3~\cite{ju2024naturalspeech} quantify prosodic similarity by comparing pitch and duration features between the prompt and output, while StyleTTS-ZS~\cite{li-etal-2025-styletts} computes Pearson correlation coefficients of acoustic features to evaluate prosodic alignment. 

Inspired by these approaches, we evaluate the prosodic similarity of MaskGCT in both English and Mandarin using the LibriTTS \cite{zen2019libritts}\footnote{\url{https://www.openslr.org/60/}} and AISHELL \cite{shi2020aishell}\footnote{\url{https://openslr.org/93/}} datasets, respectively. For each speaker, we randomly sample 50 utterances to construct the test sets. During inference, we synthesize 50 utterances per dataset by conditioning on the same target text but using different speech prompts. We then compute the acoustic-level similarity between each synthesized utterance and: (1) its paired prompt (i.e., the one used during generation), and (2) an unpaired prompt from the same speaker. This comparison allows us to quantify the extent of prompt-induced prosody similarity, which serves as an indicator of potential prosody leakage in the model.

Table~\ref{tab:prosody-leakage} presents a quantitative analysis of prompt-induced prosody similarity by comparing the acoustic differences between synthesized and prompt speech under paired and unpaired conditions. Across both LibriTTS and AISHELL-3, paired prompts consistently yield lower differences in pitch, energy, and other prosodic indicators, confirming stronger alignment in prosodic patterns. In contrast, the unpaired condition results in noticeably higher deviations, particularly in pitch mean, energy mean, and jitter, suggesting that the synthesized outputs are heavily influenced by the prosodic characteristics of the prompt.

\subsection{Seen Singer Singing Voice Synthesis}

\begin{table*}[t!]
\centering
\caption{Subjective (with 95\% confidence intervals) and objective evaluation results on the unseen test set for zero-shot SVS. Note that MCD is excluded since ground-truth alignments are unavailable in this setting.}
% \resizebox{\textwidth}{!}{
\begin{tabular}{lccc|ccc}
\toprule
\multirow{2}{*}{Model} 
& \multicolumn{3}{c}{\textbf{Subjective Evaluations}} 
& \multicolumn{3}{c}{\textbf{Objective Evaluations}} \\
\cmidrule(lr){2-4} \cmidrule(lr){5-7}
& MOS-Q $\uparrow$ & MOS-N $\uparrow$ & SMOS $\uparrow$ 
& F0-RMSE $\downarrow$ & SingMOS $\uparrow$ & SECS $\uparrow$ \\
\midrule
GT & 4.20 $\pm$ 0.12 & 4.35 $\pm$ 0.14 & 4.55 $\pm$ 0.15 & - & 4.41 & 0.932 \\
GT (Acoustic Codec) & 4.07 $\pm$ 0.18 & 4.22 $\pm$ 0.15 & 4.32 $\pm$ 0.11 & 0.015 & 4.66 & 0.921 \\
\midrule
DiffSinger & 3.75 $\pm$ 0.16 & 3.72 $\pm$ 0.18 & 3.25 $\pm$ 0.12 & 0.098 & 4.11 & 0.658 \\
VISinger2 & 3.72 $\pm$ 0.19 & 3.74 $\pm$ 0.20 & 3.31 $\pm$ 0.16 & 0.074 & 4.08 & 0.704 \\
StyleSinger & 3.48 $\pm$ 0.11 & 3.82 $\pm$ 0.18 & 3.85 $\pm$ 0.15 & 0.125 & 4.22 & 0.853 \\
SPSinger & \textbf{3.92 $\pm$ 0.15} & 4.03 $\pm$ 0.15 & 3.76 $\pm$ 0.10 & 0.065 & \textbf{4.29} & 0.844 \\
Vevo 1.5 \cite{zhang2025vevo2} & 3.72 $\pm$ 0.14 & 3.81 $\pm$ 0.12 & 4.02 $\pm$ 0.15 & 0.094 & 4.16 & 0.870 \\
\midrule
\rowcolor{gray!10}
CoMelSinger (ours) & 3.87 $\pm$ 0.18 & \textbf{4.11 $\pm$ 0.15} & \textbf{4.14 $\pm$ 0.14} & \textbf{0.048} & 4.25 & \textbf{0.897} \\
\bottomrule
\end{tabular}
% }
\label{tab:main-eval-unseen}
\end{table*}

\begin{table}[t]
\centering
\caption{Objective evaluation results on the seen test set for singing voice synthesis, comparing CoMelSinger with representative systems that employ supervised pitch information for melody control. GT denotes Ground Truth.}
\begin{tabular}{lcccc}
\toprule
Model & MCD $\downarrow$ & F0-RMSE $\downarrow$ & SingMOS $\uparrow$ \\
\midrule
CoMelSinger (ours) & 4.17 & 0.042 & \textbf{4.32} \\
\midrule
XiaoiceSing & 4.54 & 0.052 & 4.26 \\
SingAug & \textbf{4.16} & \textbf{0.035} & 3.96\\
RMSSinger & 4.33 & 0.077 & 4.15 \\
\bottomrule
\end{tabular}
\label{tab:main-eval-melody}
\end{table}

For the seen singer evaluation, we compare CoMelSinger with five strong baseline systems: DiffSinger~\cite{liu2022diffsinger}, VISinger2~\cite{zhang2022visinger2}, SPSinger~\cite{zhao2025spsinger}, StyleSinger~\cite{zhang2024stylesinger}, and Vevo 1.5 \cite{zhang2025vevo2}. To ensure a fair comparison, all models adopt HiFi-GAN~\cite{kong2020hifi} as the vocoder during both training and inference. As presented in Table~\ref{tab:main-eval-seen}, CoMelSinger achieves the highest scores across both subjective and objective metrics, demonstrating its capability to synthesize natural and expressive singing voices from seen singers.

We first note that the performance gap between the Ground Truth (GT) and GT with Acoustic Codec is minimal across all metrics, confirming that the discrete acoustic token representation introduces negligible degradation and establishing a strong upper bound for token-based SVS systems. CoMelSinger approaches this bound closely, suggesting that its improvements arise from architectural designs—particularly the disentangled modeling of melody and timbre—rather than signal-level enhancements. Among all models, CoMelSinger achieves the highest SMOS and SECS scores, indicating strong timbre consistency and accurate preservation of speaker identity, which validates the effectiveness of the in-context prompting mechanism. It also attains the lowest F0-RMSE and one of the highest SingMOS scores, reflecting precise melody reproduction and high perceptual naturalness. Furthermore, its competitive MCD score demonstrates the model’s ability to reconstruct spectral features with smooth and consistent vocal quality, confirming the effectiveness of the proposed structured melody control strategy. 

In addition, we compare CoMelSinger with several representative SVS systems that employ supervised pitch-aware conditioning under the seen-singer setting, including XiaoiceSing \cite{lu2020xiaoicesing}, SingAug \cite{DBLP:conf/interspeech/GuoSQ0J22}, and RMSSinger \cite{DBLP:conf/acl/HeLYHCLZ23}. XiaoiceSing enables precise melody control through explicit F0 modeling with residual log-F0 prediction, SingAug enhances pitch modeling in SVS by applying pitch-based and mix-up data augmentation during training, and RMSSinger achieves melody control by modeling pitch directly from realistic music scores using a diffusion-based pitch modeling approach. As shown in Table~\ref{tab:main-eval-melody}, CoMelSinger achieves competitive objective performance against these supervised systems, with comparable MCD and F0-RMSE and the highest SingMOS score. Although SingAug attains slightly lower MCD and F0-RMSE, these baseline systems rely on singer-dependent training and are not designed for zero-shot SVS, whereas CoMelSinger maintains strong performance while supporting zero-shot generalization.

\subsection{Zero-Shot Singing Voice Synthesis}
We further evaluate CoMelSinger in a zero-shot setting, where the model synthesizes singing voices from speakers not seen during training. As shown in Table~\ref{tab:main-eval-unseen}, CoMelSinger maintains strong performance across all subjective and objective metrics, exhibiting only minimal degradation compared to the seen condition. 

In contrast, baseline systems show notable declines in key timbre- and melody-related metrics such as SMOS, SECS, and F0-RMSE, underscoring their limited ability to generalize to unseen vocal identities. CoMelSinger’s robustness in zero-shot scenarios is attributed to the synergy between in-context prompting—which leverages short acoustic references to anchor timbre—and large-scale speech pretraining, which imparts transferable prosodic priors. 

Despite the inherent challenge of handling unseen timbres, CoMelSinger continues to achieve high speaker similarity while preserving accurate pitch trajectories. This balance between identity retention and melodic fidelity demonstrates the model’s strong generalization capacity. While many existing approaches face trade-offs between controllability and naturalness, CoMelSinger effectively reconciles both through its structured architecture and explicit conditioning scheme. These findings position CoMelSinger as a strong baseline for zero-shot singing voice synthesis with discrete representations. 

\subsection{Ablation Study}
\begin{table}[t]
\centering
\caption{Objective evaluation results for the ablation study. CL represents the coarse-to-fine contrastive learning strategy, where SCL and FCL respectively represents sequence and frame-level contrastive learning, SVT represents using SVT for pitch guidance. The "-w/o CL+SVT" configuration corresponds to the MaskGCT-based SVS baseline.}
% \resizebox{0.5\textwidth}{!}{
\begin{tabular}{lcccc}
\toprule
Model & MCD $\downarrow$ & F0-RMSE $\downarrow$ & SingMOS  $\uparrow$ & SECS $\uparrow$ \\
\midrule
\rowcolor{gray!10}
CoMelSinger & \textbf{4.17} & \textbf{0.042} & \textbf{4.32} & \textbf{0.912} \\
\midrule
-w/o CL & 4.91 & 0.080 & 4.12 & 0.895 \\
\hspace{1em}-w/o SCL & 4.53 & 0.062 & 4.25 & 0.900 \\
\hspace{1em}-w/o FCL & 4.82 & 0.075 & 4.18 & 0.892 \\
-w/o SVT & 5.53 & 0.194 & 3.95 & 0.883 \\
-w/o CL + SVT & 5.89 & 0.210 & 3.83 & 0.874 \\
\bottomrule
\end{tabular}
\label{tab:ablation}
% }
\end{table}

\paragraph{Component Analysis}

\begin{figure*}[t]
    \centering
    % Row 1
    \begin{minipage}[t]{0.49\textwidth}
        \centering
        \includegraphics[width=\textwidth]{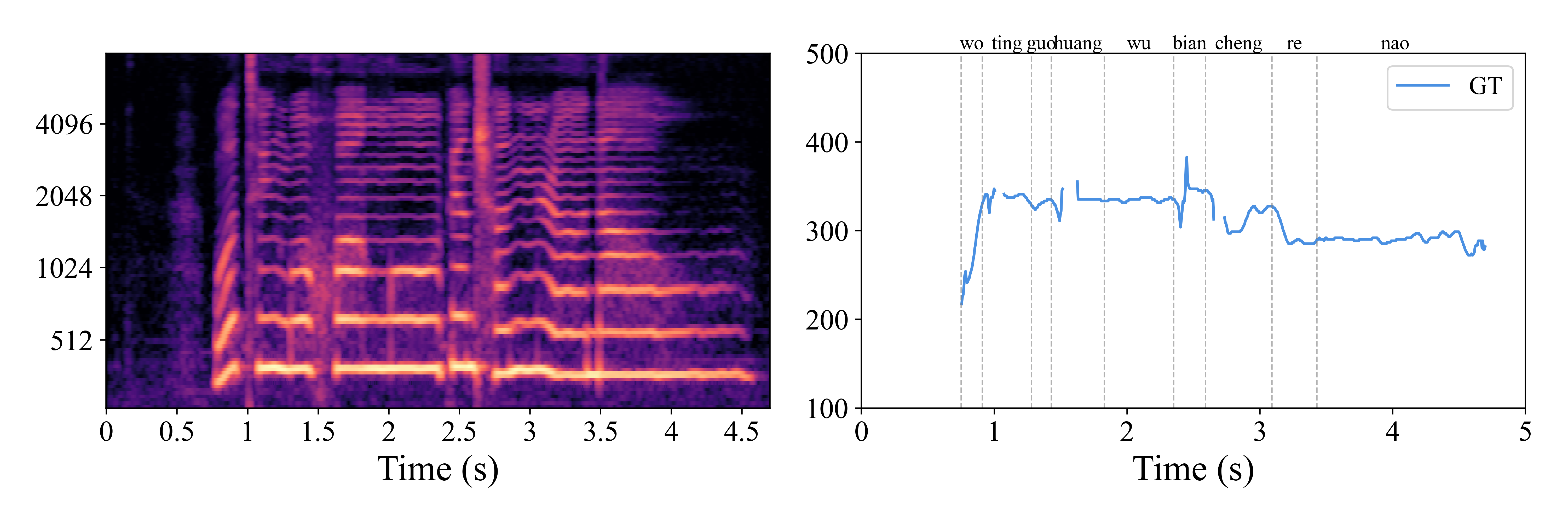}
        \par\vspace{-0.2em}
        (a) Ground Truth (GT)
    \end{minipage}
    \hfill
    \begin{minipage}[t]{0.49\textwidth}
        \centering
        \includegraphics[width=\textwidth]{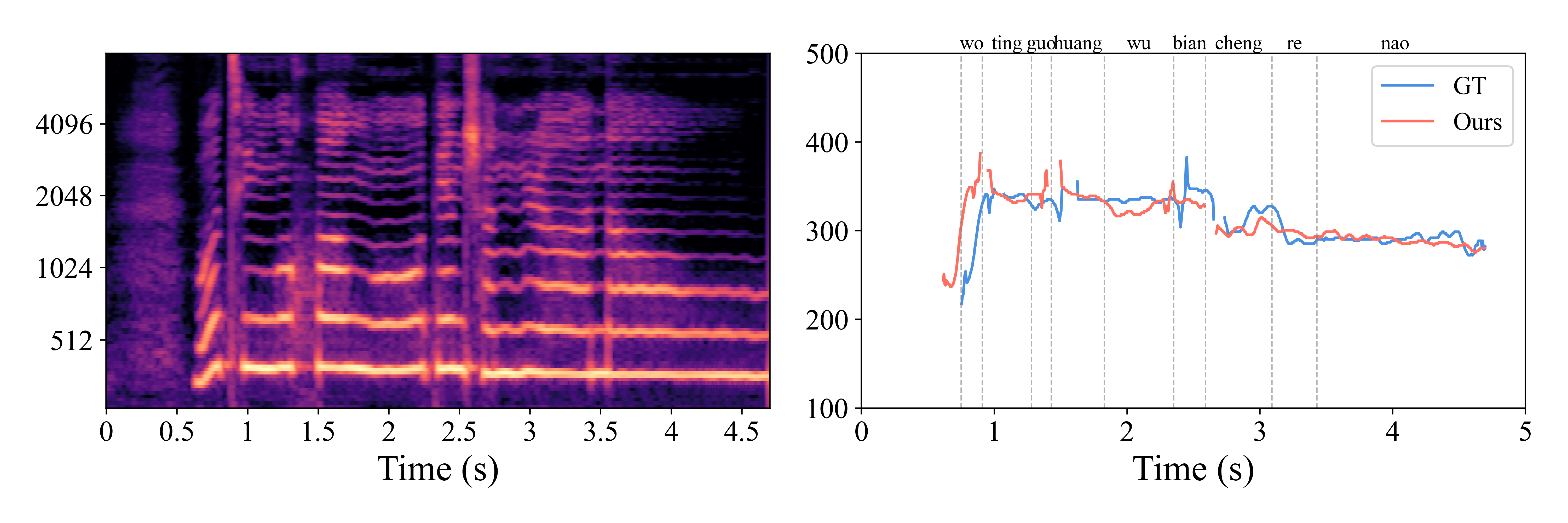}
        \par\vspace{-0.2em}
        (b) Ours
    \end{minipage}
    
    % Row 2
    \begin{minipage}[t]{0.49\textwidth}
        \centering
        \includegraphics[width=\textwidth]{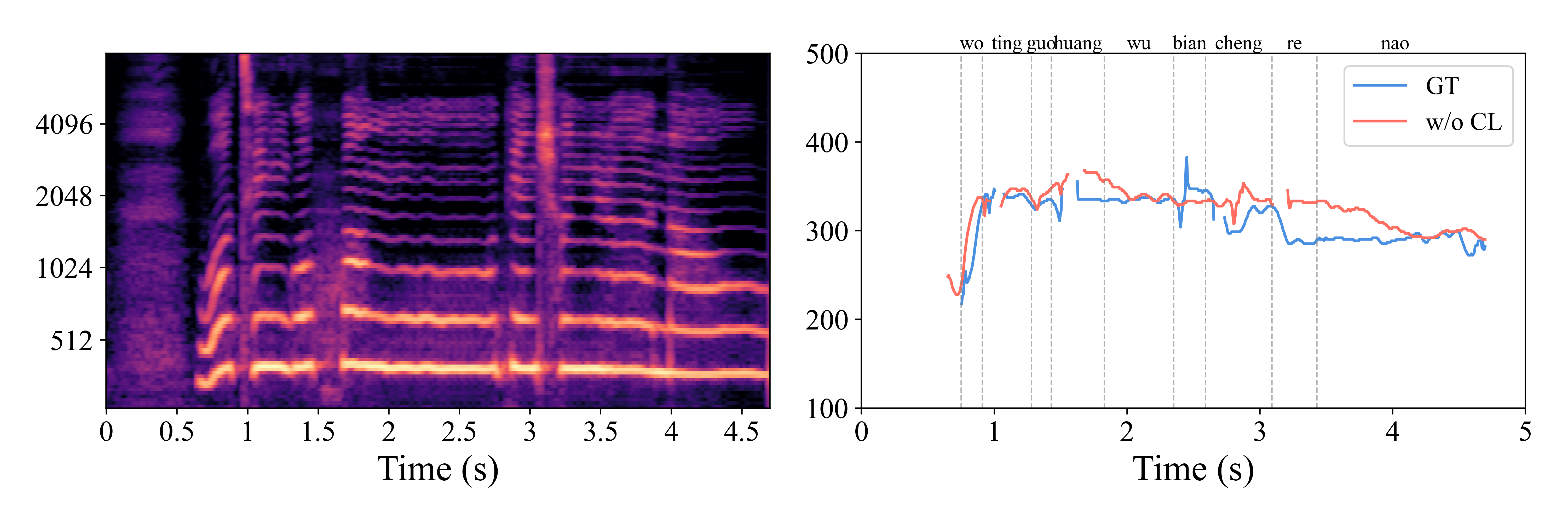}
        (c) w/o Contrastive Loss (CL)
        \par\vspace{-0.2em}
    \end{minipage}
    \hfill
    \begin{minipage}[t]{0.49\textwidth}
        \centering
        \includegraphics[width=\textwidth]{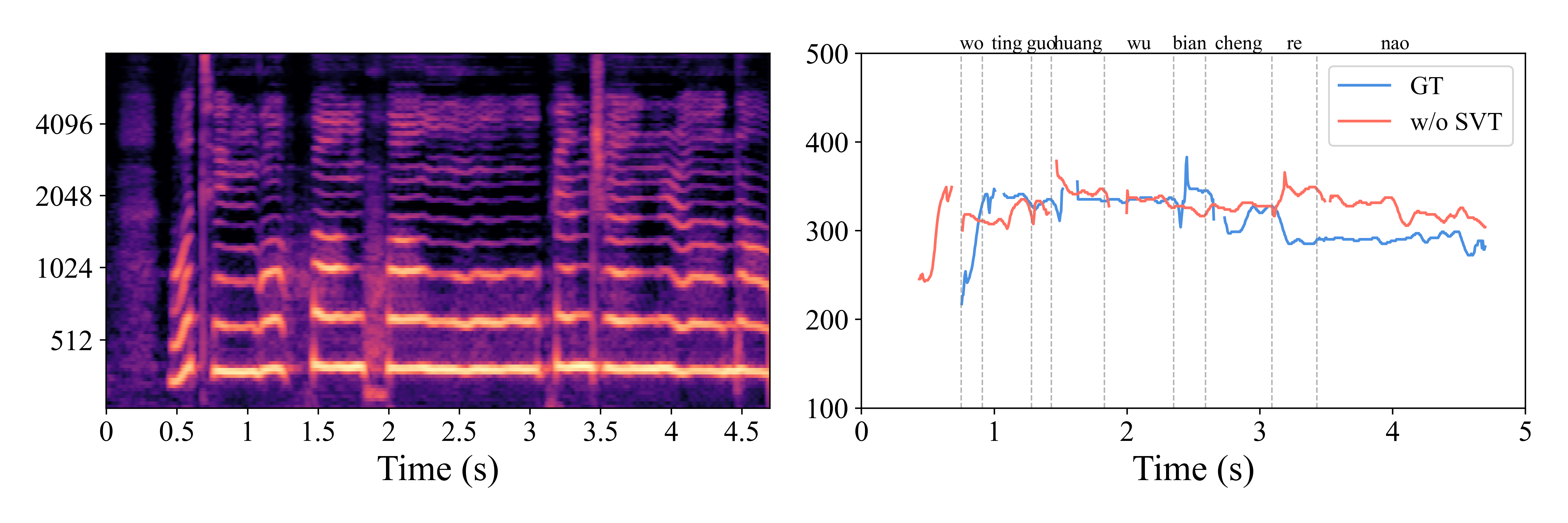}
        \par\vspace{-0.2em}
        (d) w/o SVT
    \end{minipage}
    \caption{Visualization of mel-spectrograms and pitch contours for the ground truth, the proposed model, and ablated variants. The predicted pitch trajectory (red) is overlaid with the ground-truth pitch (blue), with word-level boundaries indicated by vertical dashed lines and pinyin annotations.}
    \label{fig:mel_pitch}
\end{figure*}

To assess the contribution of each component in CoMelSinger, we perform ablation studies by systematically disabling key modules. Table~\ref{tab:ablation} reports results on four objective metrics: MCD, F0-RMSE, SingMOS, and SECS. In particular, the configuration “w/o CL + SVT” corresponds to the MaskGCT-based SVS baseline, where the S2A module is fine-tuned without explicit melody control or prosody disentanglement. Removing the entire coarse-to-fine contrastive learning (CL) framework leads to substantial degradation across all metrics, indicating reduced pitch accuracy and speaker consistency. This highlights the importance of contrastive objectives in disentangling pitch from timbre and improving input–output alignment.

To isolate the effects of each contrastive branch, we further ablate sequence contrastive learning (SCL) and frame-level contrastive learning (FCL) individually. Excluding SCL moderately affects MCD and SECS, suggesting its role in maintaining global speaker identity. In contrast, removing FCL causes a larger drop in F0-RMSE and SingMOS, confirming its effectiveness in modeling fine-grained pitch details and promoting melodic continuity. These results validate the hierarchical design of our contrastive learning framework.

We also evaluate the impact of the singing voice transcription (SVT) module, which provides auxiliary pitch supervision. Excluding SVT results in higher F0-RMSE and lower SingMOS, confirming the benefit of explicit alignment signals for structured melody control. The most severe degradation occurs when both CL and SVT are removed, indicating their complementary roles in pitch–timbre disentanglement and temporal stability. 

Figure~\ref{fig:mel_pitch} visualizes mel-spectrograms and pitch contours for the ground truth, our model, and two ablated variants. Predicted pitch trajectories (red) are overlaid with ground-truth pitch (blue), with word-level boundaries marked by dashed lines and pinyin annotations. Compared to the ablated models, CoMelSinger achieves better pitch alignment and smoother contours, illustrating the effectiveness of both CL and SVT in preserving melodic structure.

\paragraph{SVT Evaluation and Analysis}
To assess the reliability of the SVT module, we conduct an explicit frame-level evaluation using token accuracy, precision, recall, and F1 score across multiple training configurations, as summarized in Table~\ref{tab:ablation-svt}. We evaluate SVT models trained on M4Singer, Opencpop, their combination, and MIR-ST500~\cite{DBLP:conf/icassp/WangJ21}. MIR-ST500 is a large-scale singing transcription dataset comprising over 160k annotated notes from 500 pop songs. Training on the combined M4Singer and Opencpop datasets achieves the best overall performance, indicating robust pitch token prediction across diverse singing styles.

Models trained on individual datasets exhibit degraded performance, particularly on Opencpop and MIR-ST500, which can be attributed to domain mismatch and limited coverage of singing pitch patterns. We further investigate cross-dataset generalization by pretraining SVT on MIR-ST500 followed by fine-tuning on the combined dataset. Although this strategy improves precision and recall compared to training on MIR-ST500 alone, its overall performance remains inferior to training directly on singing-specific data, underscoring the importance of domain-relevant supervision for accurate pitch modeling.

\begin{table}[t]
\centering
\caption{Frame-level evaluation results of the SVT model under different training configurations. Accuracy, precision, recall, and F1 score are reported.}
\begin{tabular}{lcccc}
\toprule
Data Configuration & Accuracy & Precision & Recall & F1 \\
\midrule
M4Singer + Opencpop & \textbf{0.790} & \textbf{0.719} & \textbf{0.707} & \textbf{0.711} \\
\midrule
M4Singer & 0.749 & 0.691 & 0.674 & 0.680 \\
Opencpop & 0.707 & 0.506 & 0.485 & 0.489 \\
MIR-ST500 & 0.725 & 0.449 & 0.390 & 0.405 \\
\parbox[c]{3.2cm}{MIR-ST500 pretrain +\\ Combined finetune} & 0.648 & 0.575 & 0.542 & 0.548 \\
\bottomrule
\end{tabular}
\label{tab:ablation-svt}
\end{table}

\paragraph{Comparison of Fine-Tuning Strategies}

We evaluate six representative fine-tuning strategies on the DiffLlama backbone, each trained for 1000 epochs under identical schedules. The comparison highlights trade-offs between parameter efficiency, adaptation capacity, and overfitting risk under limited SVS data.
\begin{itemize}
    \item \textbf{FT-LoRA}: Applies LoRA to self-attention projections with $r=16$, $\alpha=32$, and dropout 0.1. Only the pitch encoder, output head, and cond module are trainable.
    
    \item \textbf{FT-LLRD}: Freezes DiffLlama and fine-tunes pitch/output/cond modules with layer-wise learning rates decayed from bottom to top: $\eta_\ell = \eta_0 \cdot \gamma^{L - 1 - \ell}$.

    \item \textbf{FT-Pitch}: Fine-tunes only the pitch encoder, output head, and cond module; the backbone remains frozen.

    \item \textbf{FT-Prefix}: Adds 20 virtual tokens to each DiffLlama layer using prefix tuning (shared across 16 layers, injected into attention and MLP). Pitch/output/cond modules are also fine-tuned.

    \item \textbf{FT-PGS}: Unfreezes two upper DiffLlama layers every 200 epochs, progressively increasing trainable capacity.

    \item \textbf{FT-Full}: Fully fine-tunes all model parameters, including the entire DiffLlama backbone.
\end{itemize}

\begin{table}[t]
\centering
\caption{Performance of various fine-tuning strategies with differing trainable parameter ratios.}
\label{tab:ft_ablation}
\resizebox{0.48\textwidth}{!}{
\begin{tabular}{lccccc}
\toprule
Method & MCD $\downarrow$ & F0-RMSE $\downarrow$ & SingMOS $\uparrow$ & SECS $\uparrow$ & Trainable (\%) \\
\midrule
FT-LoRA          & 4.26 & \textbf{0.053} & \textbf{4.34} & \textbf{0.920} & 6.51\% \\
FT-LLRD          & 4.33 & 0.069 & 4.31 & 0.894 & 100.00\% \\
FT-Pitch         & 4.47 & 0.062 & 5.95 & 0.902 & 4.46\% \\
FT-Prefix        & 4.52 & 0.059 & 6.13 & 0.911 & 5.25\% \\
FT-PGS   & \textbf{4.21} & 0.084 & 4.26 & 0.887 & 16.63\%$\sim$16.63\% \\
FT-Full & 4.41 & 0.099 & 4.13 & 0.859 & 100.00\% \\
\bottomrule
\end{tabular}
}
\end{table}

Table~\ref{tab:ft_ablation} presents a comparison of six fine-tuning strategies in terms of both objective and subjective performance, along with their respective trainable parameter ratios. FT-LoRA delivers the best overall performance, achieving the lowest F0-RMSE, highest SingMOS, and highest SECS, while updating only 4.81\% of the parameters—highlighting the effectiveness of low-rank adaptation for efficient model tuning. FT-PGS achieves the lowest MCD, suggesting enhanced spectral fidelity through gradual unfreezing, though its pitch accuracy is affected by delayed optimization of lower layers. FT-Prefix and FT-Pitch yield consistent results with minimal overhead, demonstrating the utility of lightweight adaptation modules. In contrast, FT-LLRD and FT-Full fine-tune all parameters yet underperform across most metrics, indicating that full-capacity adaptation may lead to overfitting or instability in data-scarce settings. These results underscore that parameter-efficient strategies, particularly LoRA, can match or surpass full-model fine-tuning while substantially reducing computational cost.

%% file: sections/concl.tex
In this work, we present CoMelSinger, a zero-shot singing voice synthesis framework that extends discrete token-based TTS models to support structured and controllable melody generation. Built upon the non-autoregressive MaskGCT architecture, CoMelSinger incorporates lyrics and pitch tokens as inputs, enabling fine-grained alignment between the musical score and the generated voice. To address the challenge of prosody leakage from prompt-based conditioning—an issue unique to singing synthesis—we propose a coarse-to-fine contrastive learning strategy that explicitly disentangles pitch information from the timbre prompt. Furthermore, we introduce a lightweight singing voice transcription (SVT) module to provide frame-level pitch and duration supervision, enhancing the model’s ability to follow the intended melody with precision. Extensive experiments on both seen and unseen singers demonstrate that CoMelSinger achieves strong zero-shot generalization, consistently outperforming competitive SVS baselines in pitch accuracy, timbre consistency, and subjective quality. Our results confirm that structured melody control and contrastive disentanglement are essential for scalable and expressive singing synthesis. We believe CoMelSinger opens new possibilities for discrete token-based SVS, enabling scalable and zero-shot singing generation.

%% file: Bibliography.bib
@inproceedings{zhang2024stylesinger,
  title={Stylesinger: Style transfer for out-of-domain singing voice synthesis},
  author={Zhang, Yu and Huang, Rongjie and Li, Ruiqi and He, JinZheng and Xia, Yan and Chen, Feiyang and Duan, Xinyu and Huai, Baoxing and Zhao, Zhou},
  booktitle={Proceedings of the AAAI Conference on Artificial Intelligence},
  volume={38},
  number={17},
  pages={19597--19605},
  year={2024},
doi          = {10.1609/AAAI.V38I17.29932},
}

@inproceedings{ye2023comospeech,
  title={Comospeech: One-step speech and singing voice synthesis via consistency model},
  author={Ye, Zhen and Xue, Wei and Tan, Xu and Chen, Jie and Liu, Qifeng and Guo, Yike},
  booktitle={Proceedings of the 31st ACM International Conference on Multimedia},
  pages={1831--1839},
  year={2023},
doi          = {10.1145/3581783.3612061},
}

@article{zhao2024sintechsvs,
  title={Sintechsvs: A singing technique controllable singing voice synthesis system},
  author={Zhao, Junchuan and Chetwin, Low Qi Hong and Wang, Ye},
  journal={IEEE/ACM Transactions on Audio, Speech, and Language Processing},
  volume={32},
  pages={2641--2653},
  year={2024},
  publisher={IEEE},
doi          = {10.1109/TASLP.2024.3394769},
}

@article{hwang2025hiddensinger,
  title={HiddenSinger: High-quality singing voice synthesis via neural audio codec and latent diffusion models},
  author={Hwang, Ji-Sang and Lee, Sang-Hoon and Lee, Seong-Whan},
  journal={Neural Networks},
  volume={181},
  pages={106762},
  year={2025},
  publisher={Elsevier},
doi          = {10.1016/J.NEUNET.2024.106762},
}

@inproceedings{byun2024midi,
  title={Midi-voice: Expressive zero-shot singing voice synthesis via midi-driven priors},
  author={Byun, Dong-Min and Lee, Sang-Hoon and Hwang, Ji-Sang and Lee, Seong-Whan},
  booktitle={ICASSP 2024-2024 IEEE International Conference on Acoustics, Speech and Signal Processing (ICASSP)},
  pages={12622--12626},
  year={2024},
  organization={IEEE},
doi          = {10.1109/ICASSP48485.2024.10447981},
}

@inproceedings{zhang2022visinger,
  title={Visinger: Variational inference with adversarial learning for end-to-end singing voice synthesis},
  author={Zhang, Yongmao and Cong, Jian and Xue, Heyang and Xie, Lei and Zhu, Pengcheng and Bi, Mengxiao},
  booktitle={ICASSP 2022-2022 IEEE International Conference on Acoustics, Speech and Signal Processing (ICASSP)},
  pages={7237--7241},
  year={2022},
  organization={IEEE},
doi          = {10.1109/ICASSP43922.2022.9747664},
}

@inproceedings{guo2025techsinger,
  title={Techsinger: Technique controllable multilingual singing voice synthesis via flow matching},
  author={Guo, Wenxiang and Zhang, Yu and Pan, Changhao and Huang, Rongjie and Tang, Li and Li, Ruiqi and Hong, Zhiqing and Wang, Yongqi and Zhao, Zhou},
  booktitle={Proceedings of the AAAI Conference on Artificial Intelligence},
  volume={39},
  number={22},
  pages={23978--23986},
  year={2025},
doi          = {10.1609/AAAI.V39I22.34571},
}

@inproceedings{liu2022diffsinger,
  title={Diffsinger: Singing voice synthesis via shallow diffusion mechanism},
  author={Liu, Jinglin and Li, Chengxi and Ren, Yi and Chen, Feiyang and Zhao, Zhou},
  booktitle={Proceedings of the AAAI conference on artificial intelligence},
  volume={36},
  number={10},
  pages={11020--11028},
  year={2022},
doi          = {10.1609/AAAI.V36I10.21350},
}

@article{hono2021sinsy,
  title={Sinsy: A deep neural network-based singing voice synthesis system},
  author={Hono, Yukiya and Hashimoto, Kei and Oura, Keiichiro and Nankaku, Yoshihiko and Tokuda, Keiichi},
  journal={IEEE/ACM Transactions on Audio, Speech, and Language Processing},
  volume={29},
  pages={2803--2815},
  year={2021},
  publisher={IEEE},
doi          = {10.1109/TASLP.2021.3104165},
}

@article{byun2025hierarchical,
  title={Hierarchical Diffusion Model for Zero-Shot Singing Voice Synthesis With MIDI Priors},
  author={Byun, Dong-Min and Kim, Seung-Bin and Lee, Seong-Whan},
  journal={IEEE Transactions on Audio, Speech and Language Processing},
  year={2025},
  publisher={IEEE},
doi={10.1109/TASLPRO.2025.3577324}
}

@inproceedings{dai2025everyone,
  title={Everyone-Can-Sing: Zero-Shot Singing Voice Synthesis and Conversion with Speech Reference},
  author={Dai, Shuqi and Wang, Yunyun and Dannenberg, Roger B and Jin, Zeyu},
  booktitle={ICASSP 2025-2025 IEEE International Conference on Acoustics, Speech and Signal Processing (ICASSP)},
  pages={1--5},
  year={2025},
doi          = {10.1109/ICASSP49660.2025.10889751},
  organization={IEEE}
}

@inproceedings{zhao2025spsinger,
  title={SPSinger: Multi-Singer Singing Voice Synthesis with Short Reference Prompt},
  author={Zhao, Junchuan and Low, Chetwin and Wang, Ye},
  booktitle={ICASSP 2025-2025 IEEE International Conference on Acoustics, Speech and Signal Processing (ICASSP)},
  pages={1--5},
  year={2025},
doi          = {10.1109/ICASSP49660.2025.10888907},
  organization={IEEE}
}

@article{huang2023make,
  title={Make-a-voice: Unified voice synthesis with discrete representation},
  author={Huang, Rongjie and Zhang, Chunlei and Wang, Yongqi and Yang, Dongchao and Liu, Luping and Ye, Zhenhui and Jiang, Ziyue and Weng, Chao and Zhao, Zhou and Yu, Dong},
  journal={arXiv preprint arXiv:2305.19269},
  year={2023}
}

@inproceedings{huang2021multi,
  title={Multi-singer: Fast multi-singer singing voice vocoder with a large-scale corpus},
  author={Huang, Rongjie and Chen, Feiyang and Ren, Yi and Liu, Jinglin and Cui, Chenye and Zhao, Zhou},
  booktitle={Proceedings of the 29th ACM International Conference on Multimedia},
  pages={3945--3954},
  year={2021},
doi          = {10.1145/3474085.3475437},
}

@ARTICLE{wang2023neural,
  author={Chen, Sanyuan and Wang, Chengyi and Wu, Yu and Zhang, Ziqiang and Zhou, Long and Liu, Shujie and Chen, Zhuo and Liu, Yanqing and Wang, Huaming and Li, Jinyu and He, Lei and Zhao, Sheng and Wei, Furu},
  journal={IEEE Transactions on Audio, Speech and Language Processing}, 
  title={Neural Codec Language Models are Zero-Shot Text to Speech Synthesizers}, 
  year={2025},
  volume={33},
  number={},
  pages={705-718},
  doi={10.1109/TASLPRO.2025.3530270}}

@article{du2024cosyvoice,
  title={Cosyvoice: A scalable multilingual zero-shot text-to-speech synthesizer based on supervised semantic tokens},
  author={Du, Zhihao and Chen, Qian and Zhang, Shiliang and Hu, Kai and Lu, Heng and Yang, Yexin and Hu, Hangrui and Zheng, Siqi and Gu, Yue and Ma, Ziyang and others},
  journal={arXiv preprint arXiv:2407.05407},
  year={2024}
}

@article{du2024cosyvoice2,
  title={Cosyvoice 2: Scalable streaming speech synthesis with large language models},
  author={Du, Zhihao and Wang, Yuxuan and Chen, Qian and Shi, Xian and Lv, Xiang and Zhao, Tianyu and Gao, Zhifu and Yang, Yexin and Gao, Changfeng and Wang, Hui and others},
  journal={arXiv preprint arXiv:2412.10117},
  year={2024}
}

@inproceedings{wang2024maskgct,
  author       = {Yuancheng Wang and
                  Haoyue Zhan and
                  Liwei Liu and
                  Ruihong Zeng and
                  Haotian Guo and
                  Jiachen Zheng and
                  Qiang Zhang and
                  Xueyao Zhang and
                  Shunsi Zhang and
                  Zhizheng Wu},
  title        = {MaskGCT: Zero-Shot Text-to-Speech with Masked Generative Codec Transformer},
  booktitle    = {The Thirteenth International Conference on Learning Representations,
                  {ICLR} 2025, Singapore, April 24-28, 2025},
  publisher    = {OpenReview.net},
  year         = {2025},
}

@inproceedings{ju2024naturalspeech,
  author       = {Zeqian Ju and
                  Yuancheng Wang and
                  Kai Shen and
                  Xu Tan and
                  Detai Xin and
                  Dongchao Yang and
                  Eric Liu and
                  Yichong Leng and
                  Kaitao Song and
                  Siliang Tang and
                  Zhizheng Wu and
                  Tao Qin and
                  Xiangyang Li and
                  Wei Ye and
                  Shikun Zhang and
                  Jiang Bian and
                  Lei He and
                  Jinyu Li and
                  Sheng Zhao},
  title        = {NaturalSpeech 3: Zero-Shot Speech Synthesis with Factorized Codec and Diffusion Models},
  booktitle    = {Forty-first International Conference on Machine Learning, {ICML} 2024},
  publisher    = {OpenReview.net},
  year         = {2024},
}

@article{zhang2022m4singer,
  title={M4singer: A multi-style, multi-singer and musical score provided mandarin singing corpus},
  author={Zhang, Lichao and Li, Ruiqi and Wang, Shoutong and Deng, Liqun and Liu, Jinglin and Ren, Yi and He, Jinzheng and Huang, Rongjie and Zhu, Jieming and Chen, Xiao and others},
  journal={Advances in Neural Information Processing Systems},
  volume={35},
  pages={6914--6926},
  year={2022}
}

@inproceedings{wang2022opencpop,
  author       = {Yu Wang and
                  Xinsheng Wang and
                  Pengcheng Zhu and
                  Jie Wu and
                  Hanzhao Li and
                  Heyang Xue and
                  Yongmao Zhang and
                  Lei Xie and
                  Mengxiao Bi},
  title        = {Opencpop: {A} High-Quality Open Source Chinese Popular Song Corpus
                  for Singing Voice Synthesis},
  booktitle    = {23rd Annual Conference of the International Speech Communication Association,
                  Interspeech 2022},
  pages        = {4242--4246},
  publisher    = {{ISCA}},
  year         = {2022},
  doi          = {10.21437/INTERSPEECH.2022-48},
}

@inproceedings{qian2020unsupervised,
  title={Unsupervised speech decomposition via triple information bottleneck},
  author={Qian, Kaizhi and Zhang, Yang and Chang, Shiyu and Hasegawa-Johnson, Mark and Cox, David},
  booktitle={International Conference on Machine Learning},
  pages={7836--7846},
  year={2020},
  organization={PMLR}
}

@inproceedings{tang2022avqvc,
  title={Avqvc: One-shot voice conversion by vector quantization with applying contrastive learning},
  author={Tang, Huaizhen and Zhang, Xulong and Wang, Jianzong and Cheng, Ning and Xiao, Jing},
  booktitle={ICASSP 2022-2022 IEEE International Conference on Acoustics, Speech and Signal Processing (ICASSP)},
  pages={4613--4617},
  year={2022},
doi          = {10.1109/ICASSP43922.2022.9746369},
  organization={IEEE}
}

@inproceedings{wu2024dctts,
  title={Dctts: Discrete diffusion model with contrastive learning for text-to-speech generation},
  author={Wu, Zhichao and Li, Qiulin and Liu, Sixing and Yang, Qun},
  booktitle={ICASSP 2024-2024 IEEE International Conference on Acoustics, Speech and Signal Processing (ICASSP)},
  pages={11336--11340},
  year={2024},
doi          = {10.1109/ICASSP48485.2024.10447661},
  organization={IEEE}
}

@inproceedings{zhao2022disentangling,
  title={Disentangling content and fine-grained prosody information via hybrid asr bottleneck features for voice conversion},
  author={Zhao, Xintao and Liu, Feng and Song, Changhe and Wu, Zhiyong and Kang, Shiyin and Tuo, Deyi and Meng, Helen},
  booktitle={ICASSP 2022-2022 IEEE International Conference on Acoustics, Speech and Signal Processing (ICASSP)},
  pages={7022--7026},
  year={2022},
doi          = {10.1109/ICASSP43922.2022.9747625},
  organization={IEEE}
}

@inproceedings{lian2022robust,
  title={Robust disentangled variational speech representation learning for zero-shot voice conversion},
  author={Lian, Jiachen and Zhang, Chunlei and Yu, Dong},
  booktitle={ICASSP 2022-2022 IEEE International Conference on Acoustics, Speech and Signal Processing (ICASSP)},
  pages={6572--6576},
  year={2022},
  doi          = {10.1109/ICASSP43922.2022.9747272},
  organization={IEEE}
}

@inproceedings{wu2024toksing,
  author       = {Yuning Wu and
                  Chunlei Zhang and
                  Jiatong Shi and
                  Yuxun Tang and
                  Shan Yang and
                  Qin Jin},
  title        = {TokSing: Singing Voice Synthesis based on Discrete Tokens},
  booktitle    = {25th Annual Conference of the International Speech Communication Association,
                  Interspeech 2024},
  publisher    = {{ISCA}},
  year         = {2024},
  doi          = {10.21437/INTERSPEECH.2024-2360},
}

@inproceedings{lu2020xiaoicesing,
  author       = {Peiling Lu and
                  Jie Wu and
                  Jian Luan and
                  Xu Tan and
                  Li Zhou},
  title        = {XiaoiceSing: {A} High-Quality and Integrated Singing Voice Synthesis System},
  booktitle    = {21st Annual Conference of the International Speech Communication Association,
                  Interspeech 2020},
  pages        = {1306--1310},
  publisher    = {{ISCA}},
  year         = {2020},
  doi          = {10.21437/INTERSPEECH.2020-1410},
}

@inproceedings{ren2020deepsinger,
  title={Deepsinger: Singing voice synthesis with data mined from the web},
  author={Ren, Yi and Tan, Xu and Qin, Tao and Luan, Jian and Zhao, Zhou and Liu, Tie-Yan},
  booktitle={Proceedings of the 26th ACM SIGKDD International Conference on Knowledge Discovery \& Data Mining},
  pages={1979--1989},
  doi = {10.1145/3394486.3403249},
  year={2020}
}

@inproceedings{chandna2019wgansing,
  title={Wgansing: A multi-voice singing voice synthesizer based on the wasserstein-gan},
  author={Chandna, Pritish and Blaauw, Merlijn and Bonada, Jordi and G{\'o}mez, Emilia},
  booktitle={2019 27th European signal processing conference (EUSIPCO)},
  pages={1--5},
  year={2019},
doi          = {10.23919/EUSIPCO.2019.8903099},
  organization={IEEE}
}

@inproceedings{huang2022singgan,
  title={Singgan: Generative adversarial network for high-fidelity singing voice generation},
  author={Huang, Rongjie and Cui, Chenye and Chen, Feiyang and Ren, Yi and Liu, Jinglin and Zhao, Zhou and Huai, Baoxing and Wang, Zhefeng},
  booktitle={Proceedings of the 30th ACM International Conference on Multimedia},
  pages={2525--2535},
  doi = {10.1145/3503161.3547854},
  year={2022}
}

@inproceedings{zhang2022visinger2,
  title     = {VISinger2: High-Fidelity End-to-End Singing Voice Synthesis Enhanced by Digital Signal Processing Synthesizer},
  author    = {Yongmao Zhang and Heyang Xue and Hanzhao Li and Lei Xie and Tingwei Guo and Ruixiong Zhang and Caixia Gong},
  year      = {2023},
  booktitle = {Interspeech 2023},
  pages     = {4444--4448},
  doi       = {10.21437/Interspeech.2023-391},
  issn      = {2958-1796},
}

@inproceedings{saino06_interspeech,
  author={Keijiro Saino and Heiga Zen and Yoshihiko Nankaku and Akinobu Lee and Keiichi Tokuda},
  title={{An HMM-based singing voice synthesis system}},
  year=2006,
  booktitle={Proc. Interspeech 2006},
  pages={paper 2077-Thu1BuP.7},
doi       = {10.21437/Interspeech.2006-584},
}

@inproceedings{kenmochi2007vocaloid,
  title={VOCALOID-commercial singing synthesizer based on sample concatenation},
  author={Kenmochi, Hideki and Ohshita, Hayato},
  booktitle={Interspeech},
  volume={2007},
  pages={4009--4010},
  year={2007},

}

@article{bonada2007synthesis,
  title={Synthesis of the singing voice by performance sampling and spectral models},
  author={Bonada, Jordi and Serra, Xavier},
  journal={IEEE signal processing magazine},
  volume={24},
  number={2},
  pages={67--79},
  year={2007},
  publisher={IEEE},
doi = {10.1109/MSP.2007.323266}
}

@article{hsu2021hubert,
  title={Hubert: Self-supervised speech representation learning by masked prediction of hidden units},
  author={Hsu, Wei-Ning and Bolte, Benjamin and Tsai, Yao-Hung Hubert and Lakhotia, Kushal and Salakhutdinov, Ruslan and Mohamed, Abdelrahman},
  journal={IEEE/ACM transactions on audio, speech, and language processing},
  volume={29},
  pages={3451--3460},
  year={2021},
 doi          = {10.1109/TASLP.2021.3122291},
  publisher={IEEE}
}

@article{baevski2020wav2vec,
  title={wav2vec 2.0: A framework for self-supervised learning of speech representations},
  author={Baevski, Alexei and Zhou, Yuhao and Mohamed, Abdelrahman and Auli, Michael},
  journal={Advances in neural information processing systems},
  volume={33},
  pages={12449--12460},
  year={2020}
}

@article{chen2024valle2,
  title={Vall-e 2: Neural codec language models are human parity zero-shot text to speech synthesizers},
  author={Chen, Sanyuan and Liu, Shujie and Zhou, Long and Liu, Yanqing and Tan, Xu and Li, Jinyu and Zhao, Sheng and Qian, Yao and Wei, Furu},
  journal={arXiv preprint arXiv:2406.05370},
  year={2024}
}

@article{han2024valler,
  title={Vall-e r: Robust and efficient zero-shot text-to-speech synthesis via monotonic alignment},
  author={Han, Bing and Zhou, Long and Liu, Shujie and Chen, Sanyuan and Meng, Lingwei and Qian, Yanming and Liu, Yanqing and Zhao, Sheng and Li, Jinyu and Wei, Furu},
  journal={arXiv preprint arXiv:2406.07855},
  year={2024}
}

@article{zhang2023speak,
  title={Speak foreign languages with your own voice: Cross-lingual neural codec language modeling},
  author={Zhang, Ziqiang and Zhou, Long and Wang, Chengyi and Chen, Sanyuan and Wu, Yu and Liu, Shujie and Chen, Zhuo and Liu, Yanqing and Wang, Huaming and Li, Jinyu and others},
  journal={arXiv preprint arXiv:2303.03926},
  year={2023}
}

@inproceedings{guo2024socodec,
  title={Socodec: A semantic-ordered multi-stream speech codec for efficient language model based text-to-speech synthesis},
  author={Guo, Haohan and Xie, Fenglong and Xie, Kun and Yang, Dongchao and Guo, Dake and Wu, Xixin and Meng, Helen},
  booktitle={2024 IEEE Spoken Language Technology Workshop (SLT)},
  pages={645--651},
  year={2024},
  doi          = {10.1109/SLT61566.2024.10832247},
  organization={IEEE}
}

@article{zhao2025prosody,
  title={Prosody-Adaptable Audio Codecs for Zero-Shot Voice Conversion via In-Context Learning},
  author={Zhao, Junchuan and Wang, Xintong and Wang, Ye},
  journal={arXiv preprint arXiv:2505.15402},
  year={2025}
}

@inproceedings{wang2019symmetric,
  title={Symmetric cross entropy for robust learning with noisy labels},
  author={Wang, Yisen and Ma, Xingjun and Chen, Zaiyi and Luo, Yuan and Yi, Jinfeng and Bailey, James},
  booktitle={Proceedings of the IEEE/CVF international conference on computer vision},
  pages={322--330},
  doi          = {10.1109/ICCV.2019.00041},
  year={2019}
}

@article{touvron2023llama,
  title={Llama 2: Open foundation and fine-tuned chat models},
  author={Touvron, Hugo and Martin, Louis and Stone, Kevin and Albert, Peter and Almahairi, Amjad and Babaei, Yasmine and Bashlykov, Nikolay and Batra, Soumya and Bhargava, Prajjwal and Bhosale, Shruti and others},
  journal={arXiv preprint arXiv:2307.09288},
  year={2023}
}

@inproceedings{li2024robust,
  author       = {Ruiqi Li and
                  Yu Zhang and
                  Yongqi Wang and
                  Zhiqing Hong and
                  Rongjie Huang and
                  Zhou Zhao},
  title        = {Robust Singing Voice Transcription Serves Synthesis},
  booktitle    = {Proceedings of the 62nd Annual Meeting of the Association for Computational
                  Linguistics (Volume 1: Long Papers), {ACL} 2024},
  pages        = {9751--9766},
  publisher    = {Association for Computational Linguistics},
  year         = {2024},
  doi          = {10.18653/V1/2024.ACL-LONG.526},
}

@article{chen2022wavlm,
  title={Wavlm: Large-scale self-supervised pre-training for full stack speech processing},
  author={Chen, Sanyuan and Wang, Chengyi and Chen, Zhengyang and Wu, Yu and Liu, Shujie and Chen, Zhuo and Li, Jinyu and Kanda, Naoyuki and Yoshioka, Takuya and Xiao, Xiong and others},
  journal={IEEE Journal of Selected Topics in Signal Processing},
  volume={16},
  number={6},
  pages={1505--1518},
  year={2022},
doi          = {10.1109/JSTSP.2022.3188113},
  publisher={IEEE}
}

@article{tang2024singmos,
  title={Singmos: An extensive open-source singing voice dataset for mos prediction},
  author={Tang, Yuxun and Shi, Jiatong and Wu, Yuning and Jin, Qin},
  journal={arXiv preprint arXiv:2406.10911},
  year={2024}
}

@article{borsos2023soundstorm,
  title={Soundstorm: Efficient parallel audio generation},
  author={Borsos, Zal{\'a}n and Sharifi, Matt and Vincent, Damien and Kharitonov, Eugene and Zeghidour, Neil and Tagliasacchi, Marco},
  journal={arXiv preprint arXiv:2305.09636},
  year={2023}
}

@inproceedings{nguyen2025accelerating,
  title={Accelerating codec-based speech synthesis with multi-token prediction and speculative decoding},
  author={Nguyen, Tan Dat and Kim, Ji-Hoon and Choi, Jeongsoo and Choi, Shukjae and Park, Jinseok and Lee, Younglo and Chung, Joon Son},
  booktitle={ICASSP 2025-2025 IEEE International Conference on Acoustics, Speech and Signal Processing (ICASSP)},
  pages={1--5},
  year={2025},
doi          = {10.1109/ICASSP49660.2025.10887855},
  organization={IEEE}
}

@article{pamisetty2023prosody,
  title={Prosody-TTS: An end-to-end speech synthesis system with prosody control},
  author={Pamisetty, Giridhar and Sri Rama Murty, K},
  journal={Circuits, Systems, and Signal Processing},
  volume={42},
  number={1},
  pages={361--384},
  year={2023},
doi          = {10.1007/S00034-022-02126-Z},
  publisher={Springer}
}

@inproceedings{raitio2022hierarchical,
  title={Hierarchical prosody modeling and control in non-autoregressive parallel neural TTS},
  author={Raitio, Tuomo and Li, Jiangchuan and Seshadri, Shreyas},
  booktitle={ICASSP 2022-2022 IEEE International Conference on Acoustics, Speech and Signal Processing (ICASSP)},
  pages={7587--7591},
  year={2022},
doi          = {10.1109/ICASSP43922.2022.9746253},
  organization={IEEE}
}

@inproceedings{liu2024diffstyletts,
  author       = {Jiaxuan Liu and
                  Zhaoci Liu and
                  Yajun Hu and
                  Yingying Gao and
                  Shilei Zhang and
                  Zhenhua Ling},
  title        = {DiffStyleTTS: Diffusion-based Hierarchical Prosody Modeling for Text-to-Speech
                  with Diverse and Controllable Styles},
  booktitle    = {Proceedings of the 31st International Conference on Computational
                  Linguistics, {COLING} 2025},
  pages        = {5265--5272},
  publisher    = {Association for Computational Linguistics},
  year         = {2025},
}

@inproceedings{chen2025drawspeech,
  title={DrawSpeech: Expressive Speech Synthesis Using Prosodic Sketches as Control Conditions},
  author={Chen, Weidong and Yang, Shan and Li, Guangzhi and Wu, Xixin},
  booktitle={ICASSP 2025-2025 IEEE International Conference on Acoustics, Speech and Signal Processing (ICASSP)},
  pages={1--5},
  year={2025},
  organization={IEEE}
}

@article{lee2022expressive,
  title={Expressive singing synthesis using local style token and dual-path pitch encoder},
  author={Lee, Juheon and Choi, Hyeong-Seok and Lee, Kyogu},
  journal={arXiv preprint arXiv:2204.03249},
  year={2022}
}

@inproceedings{wang2024prompt,
  author       = {Yongqi Wang and
                  Ruofan Hu and
                  Rongjie Huang and
                  Zhiqing Hong and
                  Ruiqi Li and
                  Wenrui Liu and
                  Fuming You and
                  Tao Jin and
                  Zhou Zhao},
  title        = {Prompt-Singer: Controllable Singing-Voice-Synthesis with Natural Language
                  Prompt},
  booktitle    = {Proceedings of the 2024 Conference of the North American Chapter of
                  the Association for Computational Linguistics: Human Language Technologies
                  (Volume 1: Long Papers), {NAACL} 2024},
  pages        = {4780--4794},
  publisher    = {Association for Computational Linguistics},
  year         = {2024},
  doi          = {10.18653/V1/2024.NAACL-LONG.268}
}

@inproceedings{ye-etal-2023-clapspeech,
    title = "{CLAPS}peech: Learning Prosody from Text Context with Contrastive Language-Audio Pre-Training",
    author = "Ye, Zhenhui  and
      Huang, Rongjie  and
      Ren, Yi  and
      Jiang, Ziyue  and
      Liu, Jinglin  and
      He, Jinzheng  and
      Yin, Xiang  and
      Zhao, Zhou",
    booktitle = "Proceedings of the 61st Annual Meeting of the Association for Computational Linguistics (Volume 1: Long Papers)",
    month = jul,
    year = "2023",
    publisher = "Association for Computational Linguistics",
    doi = "10.18653/v1/2023.acl-long.518",
    pages = "9317--9331",
}

@inproceedings{latif-etal-2021-controlling,
    title = "Controlling Prosody in End-to-End {TTS}: A Case Study on Contrastive Focus Generation",
    author = "Latif, Siddique  and
      Kim, Inyoung  and
      Calapodescu, Ioan  and
      Besacier, Laurent",
    booktitle = "Proceedings of the 25th Conference on Computational Natural Language Learning",
    month = nov,
    year = "2021",
    publisher = "Association for Computational Linguistics",
    doi = "10.18653/v1/2021.conll-1.42",
    pages = "544--551",
}

@inproceedings{weston2021learning,
  title={Learning de-identified representations of prosody from raw audio},
  author={Weston, Jack and Lenain, Raphael and Meepegama, Udeepa and Fristed, Emil},
  booktitle={International Conference on Machine Learning},
  pages={11134--11145},
  year={2021},
  organization={PMLR}
}

@inproceedings{xiao2024contrastive,
  title={Contrastive context-speech pretraining for expressive text-to-speech synthesis},
  author={Xiao, Yujia and Wang, Xi and Tan, Xu and He, Lei and Zhu, Xinfa and Zhao, Sheng and Lee, Tan},
  booktitle={Proceedings of the 32nd ACM International Conference on Multimedia},
  pages={2099--2107},
doi          = {10.1145/3664647.3681348},
  year={2024}
}

@inproceedings{qiang2024learning,
  title={Learning speech representation from contrastive token-acoustic pretraining},
  author={Qiang, Chunyu and Li, Hao and Tian, Yixin and Fu, Ruibo and Wang, Tao and Wang, Longbiao and Dang, Jianwu},
  booktitle={ICASSP 2024-2024 IEEE International Conference on Acoustics, Speech and Signal Processing (ICASSP)},
  pages={10196--10200},
 doi          = {10.1109/ICASSP48485.2024.10447797},
  year={2024},
  organization={IEEE}
}

@inproceedings{wang2022singing,
  title={Singing-Tacotron: Global duration control attention and dynamic filter for end-to-end singing voice synthesis},
  author={Wang, Tao and Fu, Ruibo and Yi, Jiangyan and Wen, Zhengqi and Tao, Jianhua},
  booktitle={Proceedings of the 1st International Workshop on Deepfake Detection for Audio Multimedia},
  pages={53--59},
doi          = {10.1145/3552466.3556534},
  year={2022}
}

@article{ren2019fastspeech,
  title={Fastspeech: Fast, robust and controllable text to speech},
  author={Ren, Yi and Ruan, Yangjun and Tan, Xu and Qin, Tao and Zhao, Sheng and Zhao, Zhou and Liu, Tie-Yan},
  journal={Advances in neural information processing systems},
  volume={32},
  year={2019}
}

@article{hu2022lora,
  title={Lora: Low-rank adaptation of large language models.},
  author={Hu, Edward J and Shen, Yelong and Wallis, Phillip and Allen-Zhu, Zeyuan and Li, Yuanzhi and Wang, Shean and Wang, Lu and Chen, Weizhu and others},
  journal={ICLR},
  volume={1},
  number={2},
  pages={3},
  year={2022}
}

@article{kong2020hifi,
  title={Hifi-gan: Generative adversarial networks for efficient and high fidelity speech synthesis},
  author={Kong, Jungil and Kim, Jaehyeon and Bae, Jaekyoung},
  journal={Advances in neural information processing systems},
  volume={33},
  pages={17022--17033},
  year={2020}
}

@inproceedings{shen2023naturalspeech,
  author       = {Kai Shen and
                  Zeqian Ju and
                  Xu Tan and
                  Eric Liu and
                  Yichong Leng and
                  Lei He and
                  Tao Qin and
                  Sheng Zhao and
                  Jiang Bian},
  title        = {NaturalSpeech 2: Latent Diffusion Models are Natural and Zero-Shot
                  Speech and Singing Synthesizers},
  booktitle    = {The Twelfth International Conference on Learning Representations,
                  {ICLR} 2024},
  publisher    = {OpenReview.net},
  year         = {2024},
}

@inproceedings{li-etal-2025-styletts,
    title = "{S}tyle{TTS}-{ZS}: Efficient High-Quality Zero-Shot Text-to-Speech Synthesis with Distilled Time-Varying Style Diffusion",
    author = "Li, Yinghao Aaron  and
      Jiang, Xilin  and
      Han, Cong  and
      Mesgarani, Nima",
    booktitle = "Proceedings of the 2025 Conference of the Nations of the Americas Chapter of the Association for Computational Linguistics: Human Language Technologies (Volume 1: Long Papers)",
    month = apr,
    year = "2025",
    publisher = "Association for Computational Linguistics",
    doi = "10.18653/v1/2025.naacl-long.242",
    pages = "4725--4744",
}

@inproceedings{zen2019libritts,
  author       = {Heiga Zen and
                  Viet Dang and
                  Rob Clark and
                  Yu Zhang and
                  Ron J. Weiss and
                  Ye Jia and
                  Zhifeng Chen and
                  Yonghui Wu},
  title        = {LibriTTS: {A} Corpus Derived from LibriSpeech for Text-to-Speech},
  booktitle    = {20th Annual Conference of the International Speech Communication Association,
                  Interspeech 2019},
  pages        = {1526--1530},
  publisher    = {{ISCA}},
  year         = {2019},
  doi          = {10.21437/INTERSPEECH.2019-2441},
}

@article{shi2020aishell,
  title={Aishell-3: A multi-speaker mandarin tts corpus and the baselines},
  author={Shi, Yao and Bu, Hui and Xu, Xin and Zhang, Shaoji and Li, Ming},
  journal={arXiv preprint arXiv:2010.11567},
  year={2020}
}

@inproceedings{dai2023singstyle111,
  author       = {Shuqi Dai and
                  Yuxuan Wu and
                  Siqi Chen and
                  Roy Huang and
                  Roger B. Dannenberg},
  title        = {SingStyle111: {A} Multilingual Singing Dataset With Style Transfer},
  booktitle    = {Proceedings of the 24th International Society for Music Information
                  Retrieval Conference, {ISMIR} 2023},
  pages        = {765--773},
  year         = {2023},
  doi          = {10.5281/ZENODO.10265401}
}

@inproceedings{kahn2020libri,
  title={Libri-light: A benchmark for asr with limited or no supervision},
  author={Kahn, Jacob and Riviere, Morgane and Zheng, Weiyi and Kharitonov, Evgeny and Xu, Qiantong and Mazar{\'e}, Pierre-Emmanuel and Karadayi, Julien and Liptchinsky, Vitaliy and Collobert, Ronan and Fuegen, Christian and others},
  booktitle    = {2020 {IEEE} International Conference on Acoustics, Speech and Signal
                  Processing, {ICASSP} 2020},
  pages={7669--7673},
  year={2020},
  organization={IEEE}
}

@inproceedings{he2024emilia,
  title={Emilia: An extensive, multilingual, and diverse speech dataset for large-scale speech generation},
  author={He, Haorui and Shang, Zengqiang and Wang, Chaoren and Li, Xuyuan and Gu, Yicheng and Hua, Hua and Liu, Liwei and Yang, Chen and Li, Jiaqi and Shi, Peiyang and others},
  booktitle    = {{IEEE} Spoken Language Technology Workshop, {SLT} 2024},
  pages={885--890},
  year={2024},
  organization={IEEE}
}

@article{zhang2025vevo2,
  title={Vevo2: Bridging controllable speech and singing voice generation via unified prosody learning},
  author={Zhang, Xueyao and Zhang, Junan and Wang, Yuancheng and Wang, Chaoren and Chen, Yuanzhe and Jia, Dongya and Chen, Zhuo and Wu, Zhizheng},
  journal={arXiv preprint arXiv:2508.16332},
  year={2025}
}

@inproceedings{DBLP:conf/interspeech/GuoSQ0J22,
  author       = {Shuai Guo and
                  Jiatong Shi and
                  Tao Qian and
                  Shinji Watanabe and
                  Qin Jin},
  title        = {SingAug: Data Augmentation for Singing Voice Synthesis with Cycle-consistent
                  Training Strategy},
  booktitle    = {23rd Annual Conference of the International Speech Communication Association,
                  Interspeech 2022, Incheon, Korea, September 18-22, 2022},
  pages        = {4272--4276},
  publisher    = {{ISCA}},
  year         = {2022},
  doi          = {10.21437/INTERSPEECH.2022-978},
}

@inproceedings{DBLP:conf/acl/HeLYHCLZ23,
  author       = {Jinzheng He and
                  Jinglin Liu and
                  Zhenhui Ye and
                  Rongjie Huang and
                  Chenye Cui and
                  Huadai Liu and
                  Zhou Zhao},
  title        = {RMSSinger: Realistic-Music-Score based Singing Voice Synthesis},
  booktitle    = {Findings of the Association for Computational Linguistics: {ACL} 2023,
                  Toronto, Canada, July 9-14, 2023},
  pages        = {236--248},
  publisher    = {Association for Computational Linguistics},
  year         = {2023},
  doi          = {10.18653/V1/2023.FINDINGS-ACL.16},
}

@inproceedings{DBLP:conf/icassp/WangJ21,
  author       = {Jun{-}You Wang and
                  Jyh{-}Shing Roger Jang},
  title        = {On the Preparation and Validation of a Large-Scale Dataset of Singing
                  Transcription},
  booktitle    = {{IEEE} International Conference on Acoustics, Speech and Signal Processing,
                  {ICASSP} 2021, Toronto, ON, Canada, June 6-11, 2021},
  pages        = {276--280},
  publisher    = {{IEEE}},
  year         = {2021},
  doi          = {10.1109/ICASSP39728.2021.9414601},
}
